\documentclass[trackchanges]{aastex63}

\received{June 5, 2020}
\revised{July 15, 2020}
\accepted{July 25, 2020}
\shorttitle{Anisotropy of Long-period Comets}
\shortauthors{Higuchi}
\begin{document}

\title{Anisotropy of Long-period Comets Explained by Their Formation Process}

\correspondingauthor{Arika Higuchi}
%\email{higuchi.arika@nao.ac.jp, higuchia@gmail.com}
\email{higuchiarika@med.uoeh-u.ac.jp}

%\author[0000-0002-1505-8083]{Arika Higuchi}
\author{Arika Higuchi}
\affiliation{Department of Basic Sciences, University of Occupational and Environmental Health, Japan,\\
  1-1 Isyogaoka, Yahata-nishi, Kitakyusyu,\\
  Fukuoka 807-8555, Japan}
\affiliation{(before 2020/6/30) RISE Project, National Astronomical Observatory of Japan \\
  2-21-1 Osawa, Mitaka,
  Tokyo 181-8588, Japan}



\begin{abstract}
  Long-period comets coming from the Oort cloud are thought to be
  planetesimals formed in the planetary region on the ecliptic plane.
  We have investigated the orbital evolution of these bodies
    due to the Galactic tide.
  We extended \citet{2007AJ....134.1693H} and derived the
  analytical solutions to the Galactic longitude and latitude of
  the direction of aphelion, $L$ and $B$.
  Using the analytical solutions,
  we show that the ratio of the periods of the evolution of
  $L$ and $B$ is very close to either 2 or $\infty$
  for initial eccentricities $e_i\simeq 1$,
  as is true for the Oort cloud comets.
  From the relation between $L$ and $B$,
  we predict that Oort cloud comets returning to the planetary region
  are concentrated on the ecliptic plane
  and a second plane, which we call the "empty ecliptic."
  This consists in a rotation %of 180$^\circ$
  of the ecliptic around the Galactic pole by 180$^\circ$.
  Our numerical integrations confirm
  that the radial component of the Galactic tide,
  which is neglected in the derivation of the analytical solutions, 
  is not strong enough to break the relation between
  $L$ and $B$ derived analytically.
  Brief examination of observational data shows
  that there are concentrations 
  near both the ecliptic and the empty ecliptic.
  We also show that the anomalies of the distribution of
  $B$ of long-period comets mentioned by several authors
  are explained by the concentrations on the two planes
  more consistently than by previous explanations.
\end{abstract}

\keywords{comets}
  

\section{Introduction}
The tidal force from the Galactic disk
is the dominant external force in the evolution of the bodies in
the Oort cloud \citep[e.g.,][]{2004come.book..153D}.
The vertical component (i.e., perpendicular to the Galactic plane) of the Galactic tide plays
the most important role in
the formation of the Oort cloud and the production
of long-period comets from it 
\citep[e.g.,][]{1985Icar...61...60H, 1986EM&P...36..263B, 1986Icar...65...13H}.
The Galactic potential is often approximated as axisymmetric
by neglecting the radial component.

The vertical component of the tide acts on comets in the Oort cloud 
like a secular perturbation from a planet does on asteroids, and
it drives the von Zeipel-Lidov-Kozai mechanism
\citep{1910AN....183..345V, 1962AJ.....67..591K,1962P&SS....9..719L,2019MEEP....7....1I}
in the orbital evolution, as first shown in \citet{1986Icar...65...13H}.
The time-averaged disturbing function that arises from
the vertical component of the Galactic tide
is obtained by averaging the Galactic potential over one orbital
period of the comet \citep{1986Icar...65...13H}.
By substituting the time-averaged disturbing function
into the variational equations of orbital elements
\citep[e.g.,][]{1999ssd..book.....M},
time variations of the secular orbital elements are obtained
\citep[e.g.,][]{1989Icar...82..389M,
  1992CeMDA..54...13M,
  2001MNRAS.324.1109B,
  2005MNRAS.364.1222B,
  2006Icar..184...59B,
  2007AJ....134.1693H}.
\citet{2007AJ....134.1693H} applied the solutions to
examine the formation of the Oort cloud from planetesimals with
large semimajor axes initially on the ecliptic plane.

The analytical solutions to the orbital elements
presented by the above authors 
are useful for understanding the evolution of the Oort cloud
and the overall behavior of the distribution
of the comets generated by the Galactic tide.
However, the solutions are not so useful for the discussion about
observed long-period comets returning to the planetary
region for the following two reasons.
First, the time variation of the longitude of the ascending node in the
Galactic coordinates, $d\Omega/dt$, becomes large as the eccentricity
$e$ approaches 1 \citep{2007AJ....134.1693H}.
This means that $\Omega$ of a long-period comet
and the inclination with respect to the ecliptic plane $i_{\rm E}$,
which is a function of $\Omega$ and the inclination with
respect to the Galactic plane $i$,
are drastically changing with the perihelion distance $q$
when it is in the observable region (i.e., $e\simeq 1$).
Consequently, there is no firm relation between
the initial orbital elements
and the observed orbital elements in the planetary region.
Second, the angular momentum of a comet with $e\simeq 1$
is quite small and it is easily changed
by perturbations from passing stars and/or
the radial component of the Galactic tide,
both of which are neglected in the derivation of the analytical solutions
\citep[e.g.,][]{1999Icar..141..354M, 2007AJ....134.1693H}.
The conservation of the vertical component of the
angular momentum, which is defined as $j=\sqrt{1-e^2}\cos i$,
is crucial in order to derive the analytical solution to the inclination
at small $q$.
Therefore, the accurate prediction of $i$ at $e\simeq 1$ is difficult.
For the above two reasons, the analytical solutions to the orbital
elements, especially to $\Omega$, $i$,
and $i_{\rm E}$, are not so useful for describing the orbits
of long-period comets.

Besides $\Omega$, $i$, and the argument of perihelion, $\omega$,
in the Galactic coordinates,
the Galactic longitude and latitude of the direction of aphelion,
$L$ and $B$ (or those of perihelion, $l$ and $b$),
are also used to evaluate the distribution of observed long-period comets.
Many authors have pointed out anomalies in the distributions of $L$ and $B$.
For example,
\citet{1984A&A...141...94L} and \citet{1987A&A...187..913D} 
found depletions around $b=0$ and $b=\pm 90^\circ$.
\citet{1987A&A...187..913D} explained that
the depletions are the result of the strength of the Galactic tide,
which is minimum for $b=0$ and $b=\pm90^\circ$.
\citet{1996ApJ...472L..41M} evaluated the 
effect of the radial component of the Galactic tide in
the distributions of $l$ and $b$ of long-period comets.
\citet{1983PNAS...80.5151B},
\citet{1984A&A...141...94L}, and \citet{1986gss..conf..173D}
investigated aphelion clustering on the $L-B$ plane and
\citet{1999Icar..141..354M} identified
an anomalously overpopulated ``great circle''
as two peaks centered on $L=135^\circ$ and $315^\circ$.
These concentrations of the aphelia were explained
by introducing a hypothetical perturber
that encountered the solar system.

The above investigations of the distribution of aphelia
are made on the assumption that the Oort cloud,
which stores the long-period comets,
has an isotropic distribution of the comets.
However, based on the standard formation scenario,
the Oort cloud comets are planetesimals formed in
the protoplanetary disk and
initially on the ecliptic plane
with the perihelion distances near the giant planets
\citep[e.g.,][]{2004come.book..153D}.
The role of stars in the evolution of the Oort cloud
has been examined by many authors
\citep[e.g.,][]{2002A&A...396..283D,2011Icar..214..334F}.
They showed that passing stars
act like random noise on the distribution
of comets in the Oort cloud.
As long as the Oort cloud is not completely
destroyed by close stellar encounters,
the memory of the initial distribution can be found
as anisotropies in the present distribution,
which can be explained without
assuming any hypothetical perturber.

In this paper,
we investigate the evolution of the aphelia of comets initially
on the ecliptic plane
under the axisymmetric approximation of the Galactic tide
with the same procedure as in \citet{2007AJ....134.1693H}.
Using the analytical solutions,
we predict the distribution of long-period comets on the $L-B$ plane.
The solutions to $L$, $B$, and other orbital elements are derived
in Section 2.
In Section 3, the analytic solutions are evaluated by
comparisons with numerical integrations of the equation of motion
that take into account not only the vertical but also
the radial component of the Galactic tide.
In Section 4, we approximate the analytical solutions
for the special case of Oort cloud comets and
propose the concentration of comets on the ecliptic plane
and the ``empty ecliptic'' plane,
which is defined as a plane formed by a rotation
of the ecliptic around the Galactic pole by 180$^\circ$.
In Section 5, the distribution of observed small bodies
is briefly examined to find the concentrations on the ecliptic and
the empty ecliptic in the $L-\sin B$ plane.
Section 6 is devoted to a summary and discussion.

%\newpage
\section{Analytical expression for orbital evolution}\label{ss:ana}
In this section, we derive
the Galactic longitude and latitude of the direction of the aphelion,
$L$ and $B$, respectively, and their time variations and solutions.
Time variations and solutions for the eccentricity $e$ and
the longitude of the ascending node $\Omega$
are also shown but we use slightly different expressions from
\citet{2007AJ....134.1693H} for the purpose of this paper.
The orbital elements are given in Galactic coordinates
except for the ecliptic inclination $i_{\rm E}$. 

\subsection{The Galactic longitude $L$ and latitude $B$}
Using the orbital elements,
the unit vector of the direction of aphelion
in the Galactic coordinates ${\bf r}_{\rm Q}$
is written as
\begin{eqnarray}
  {\bf r}_{\rm Q} =
  \left(
  \begin{array}{c}
    Q_x\\
    Q_y\\
    Q_z
  \end{array}
  \right)
  =
  \left(   
  \begin{array}{l}
    -\cos\omega\cos\Omega+\sin\omega\sin\Omega\cos i\\
    -\cos\omega\sin\Omega-\sin\omega\cos\Omega\cos i\\
    -\sin\omega\sin i
  \end{array}      
  \right)
\end{eqnarray}
Then $L$ and $\sin B$ are written as
\begin{equation}
  L={\rm atan}\left(\frac{Q_y}{Q_x}\right)=\Omega + \theta,
  \label{eq:elu}
\end{equation}
\begin{equation}
  \sin B=Q_z=-\sin\omega\sin i,
  \label{eq:sinb}
\end{equation}
where
\begin{eqnarray}
  \theta
  &=&\left\{
  \begin{array}{cc}
    {\rm atan}\left(\frac{\sin\omega\cos i}{\cos\omega}\right)
    &\;\;{\rm for}\;\;\cos\omega<0
    \\
    \pi+{\rm atan}\left(\frac{\sin\omega\cos i}{\cos\omega}\right)
    &\;\;{\rm for}\;\;\cos\omega>0.
  \end{array}
  \right.
  \label{eq:theta}
\end{eqnarray}

\subsection{Conserved quantities}
Assume that the Galactic tide is much smaller than the solar gravity.
The time-averaged Hamiltonian of a body moving under the
approximated Galactic potential is given as
\begin{equation}
  \langle H\rangle = -\frac{GM_\odot}{2a}-R,
  \label{eq:h}
\end{equation}
where $G$ is the gravitational constant, $M_\odot$ is the solar mass,
$a$ is the semimajor axis of the body,
and $R$ is the disturbing function
\begin{equation}
  R = -\frac{\nu_0^2}{4}a^2\sin^2 i\left(1-e^2+5e^2\sin^2\omega\right),
  \label{eq:df}
\end{equation}
where $\nu_0=\sqrt{4\pi G\rho}$ is the vertical frequency
and $\rho$ is the total density in the solar neighborhood
\citep[e.g.,][]{1986Icar...65...13H}.
From Equation (\ref{eq:df}) and Lagrange's planetary equation for $da/dt$,
we know $a$ is constant.
Then we introduce a new simplified Hamiltonian:
\begin{equation}
  c = \sin^2 i\left(1-e^2+5e^2\sin^2\omega\right).
  \label{eq:c}
\end{equation}
The simplified $z$-component of the angular momentum, which is
a conserved quantity under the axisymmetric approximation
of the potential, is written as
\begin{equation}
  j=\sqrt{1-e^2}\cos i.
  \label{eq:j}
\end{equation}

Substituting Equation (\ref{eq:j}) into Equation (\ref{eq:c}),
%$i$ is vanished and then
%Equation (\ref{eq:c}) can be used to
%draw equi-Hamiltonian curves on the $\omega-e$ plane
%for given $c$ and $j$.
one can draw equi-Hamiltonian curves on the $\omega-e$ plane
for given $c$ and $j$ using Equation (\ref{eq:c}).
From the Hamiltonian curves, we can learn the overall behavior
of $\omega$ and $e$ without solving the equation of motion.
For some cases,
equi-Hamiltonian curves circulate with $\omega$
and for other cases they librate around
$\omega=90^\circ$ or 270$^\circ$.
This libration is essentially 
the von Zeipel-Lidov-Kozai mechanism
\citep{1910AN....183..345V, 1962AJ.....67..591K,1962P&SS....9..719L,2019MEEP....7....1I}.
The condition for circulation is to have a solution to $e$
for $\omega=0$, i.e.,
\begin{eqnarray}
  \begin{array}{cccl}
  c+j^2&<&1 & \;\;{\rm (circulation)}\\
  c+j^2&>&1 & \;\;{\rm (libration)}
  \end{array}
  \label{eq:kozai}
\end{eqnarray}
and $c+j^2=1$ gives the separatrix \citep[e.g.,][]{2007AJ....134.1693H}.
This leads to the necessary condition on $i$ for libration, $\sin i>\sqrt{1/5}$.

Substituting Equations (\ref{eq:sinb}) and (\ref{eq:j}) into
Equation (\ref{eq:c}),
the Hamiltonian is given with $B$ instead of $i$ and $\omega$,
\begin{equation}
  c = 1-e^2-j^2+5e^2\sin^2 B.
  \label{eq:cb}
\end{equation}
The separatrix with Equation (\ref{eq:cb}) is written as
\begin{equation}
  c + j^2 = 1-e^2\left(1-5\sin^2 B\right)=1.
  \label{eq:sepa}
\end{equation}
For $e^2>0$,  
the sufficient condition on $B$ for libration is given as
\begin{equation}
  \sin |B|>\sqrt{\frac{1}{5}}.
  \label{eq:conb}
\end{equation}
\citet{1989Icar...82..389M} defined the value $B={\rm asin}\sqrt{1/5}\simeq \pm 26.6^\circ$
as a barrier that the latitude of perihelion cannot migrate across.

\subsection{Time variations}
Substituting Equation (\ref{eq:df}) into
Lagrange's planetary equations \citep{1999ssd..book.....M},
we obtain the time variations of $e$, $i$, $\Omega$, and $\varpi$ as
\begin{eqnarray}
  \frac{de}{dt}
  &=&
  \frac{5\nu_0^2}{2n}e\sqrt{1-e^2}\sin^2i\sin\omega\cos\omega,
  \label{eq:dedt}
  \\
  \frac{di}{dt} &=&
  -\frac{5\nu_0^2}{2n}\frac{e^2}{\sqrt{1-e^2}}\sin i\cos i\sin\omega\cos\omega,
  \label{eq:didt}
  \\
  \frac{d\Omega}{dt} &=&
  -\frac{\nu_0^2}{2n}\frac{\cos i}{\sqrt{1-e^2}}(1-e^2+5e^2\sin^2\omega),
  \label{eq:dlodt}
  \\
  \frac{d\varpi}{dt}
  &=&
  \frac{\nu_0^2}{2n}\frac{1}{\sqrt{1-e^2}}
  \left(
  1-e^2+5e^2\sin^2\omega-5\sin^2\omega\sin^2 i
  \right)  
  +\frac{d\Omega}{dt},
  \label{eq:dvpdt}
\end{eqnarray}
where $n=a^{-3/2}$ is the mean motion
and $\varpi=\omega+\Omega$ is the longitude of the pericenter.
Note that the short-period terms arising from the variation of
the mean longitude are dropped.
From the definition $\varpi=\omega+\Omega$,
\begin{eqnarray}
  \frac{d\omega}{dt} &=& \frac{d\varpi}{dt} -\frac{d\Omega}{dt}
  \nonumber\\
  &=&
  \frac{\nu_0^2}{2n}\frac{1}{\sqrt{1-e^2}}
  \left(
    1-e^2+5e^2\sin^2\omega-5\sin^2\omega\sin^2 i.
    \right)
  \label{eq:dodt}
\end{eqnarray}
From Equation (\ref{eq:elu}), we have
\begin{eqnarray}
  \frac{d L}{dt}&=&\frac{d\Omega}{dt}+\frac{d\theta}{dt}.
  \label{eq:dldt}
\end{eqnarray}
From Equation (\ref{eq:theta}), 
we have
\begin{eqnarray}
  \frac{d\theta}{dt}
  &=&
  \left(1+\tan^2\theta\right)^{-1}\left(
  \cos^{-2}\omega\frac{d\omega}{dt}\cos i
  -\tan\omega\sin i\frac{di}{dt}\right).
  \label{eq:dtdt}
\end{eqnarray}
Substituting Equations (\ref{eq:theta}), (\ref{eq:sinb}), (\ref{eq:didt}), and (\ref{eq:dodt})
into Equation (\ref{eq:dtdt}), we have
\begin{eqnarray}
  \frac{d\theta}{dt}
  &=&  -\frac{d\Omega}{dt}
  -\frac{2\nu_0^2}{n}\sqrt{1-e^2}\cos i\tan^2 B.
  \label{eq:dtdt2}
\end{eqnarray}
Substituting Equation (\ref{eq:dtdt2}) into Equation (\ref{eq:dldt}),
we have
\begin{eqnarray}
  \frac{dL}{dt}
  &=&
  -\frac{2\nu_0^2}{n}\sqrt{1-e^2}\cos i\tan^2 B.
  \label{eq:dldt2}
\end{eqnarray}

\subsection{Solutions}
\subsubsection{Solutions to $e$, $i$, $\omega$, and $B$}

Introducing $\chi=1-e^2$, Equation (\ref{eq:dedt}) can be rewritten as
\begin{eqnarray}
  \frac{d\chi}{dt}& = &\frac{d\chi}{de}\frac{de}{dt}\nonumber\\
  &=& -\frac{5\nu_0^2}{n}\chi^{-\frac{1}{2}}(1-\chi)(\chi-j^2)\sin\omega\cos\omega.
  \label{eq:dxdt}
\end{eqnarray}
Using Equations (\ref{eq:c}) and (\ref{eq:j}),
$\omega$ can be rewritten as
\begin{eqnarray}
  \sin^2\omega&=&
  \frac{\chi(c+j^2-\chi)}{5(1-\chi)\left(\chi-j^2\right)}.
  \label{eq:sin2o}
\end{eqnarray}
Substituting $\sin\omega$ and $\cos\omega$ from Equation (\ref{eq:sin2o})
into Equation (\ref{eq:dxdt}),
\begin{eqnarray}
  \frac{d\chi}{dt}
  &=&-A_1\sqrt{(\chi_0^*-\chi)(\chi_2^*-\chi)(\chi-\chi_1^*)},
  \label{eq:dxdt2}
\end{eqnarray}
where
\begin{equation}
  A_1=\frac{2\nu_0^2}{n},
  \label{eq:A_1}
\end{equation}
\begin{equation}
  \chi_0^*=c+j^2,
  \label{eq:chi0}
\end{equation}
\begin{eqnarray}
  Q(\chi)
  &=&-4\chi^2+\left(5+4j^2-c\right)\chi-5j^2
  \nonumber\\
  &\equiv&4(\chi-\chi_1^*)(\chi_2^*-\chi) \;\;\;\;(\chi_1^*<\chi_2^*).
  \label{eq:qq}
\end{eqnarray}
The solution to Equation (\ref{eq:dxdt2}) is expressed using a Jacobian
elliptic function, sn \citep{yellow},
\begin{eqnarray}
  \chi&=&(\alpha_1-\alpha_0){\rm sn}^2u+\alpha_0,
  \label{eq:chi}
\end{eqnarray}
where we define
$\alpha_0 = min\{\chi_0^*,\chi_1^*,\chi_2^*\}$,
$\alpha_1 = med\{\chi_0^*,\chi_1^*,\chi_2^*\}$,
and $\alpha_2 = max\{\chi_0^*,\chi_1^*,\chi_2^*\}$,
\begin{eqnarray}
  u&=&\frac{A_1}{g}t+u_0,
  \label{eq:u}
  \\
  g&=&\frac{2}{\sqrt{\alpha_2-\alpha_0}},
  \label{eq:g}
  \\
  u_0&=&\pm F(\phi_0,k),
  \label{eq:u0}
\end{eqnarray}
where $F$ is a normal elliptic integral of the first kind with modulus $k$ and the amplitude
$\phi_0$,
\begin{eqnarray}  
  k^2&=&\frac{\alpha_1-\alpha_0}{\alpha_2-\alpha_0}
  \label{eq:k2}
  \\
  \sin\phi_0 &=&\sqrt{\frac{\chi_0^*-\alpha_0}{\alpha_1-\alpha_0}},
  \label{eq:phi}
\end{eqnarray}
The sign of $F(\phi_0,k)$ in Equation (\ref{eq:u0})
depends on $\omega_i$, the value of $\omega$ at $t=0$;
it is negative for $\sin(2\omega_i)>0$ and positive for
$\sin(2\omega_i)<0$.

\begin{figure}[hbtp]
  \begin{center}
    \resizebox{8cm}{!}{\includegraphics{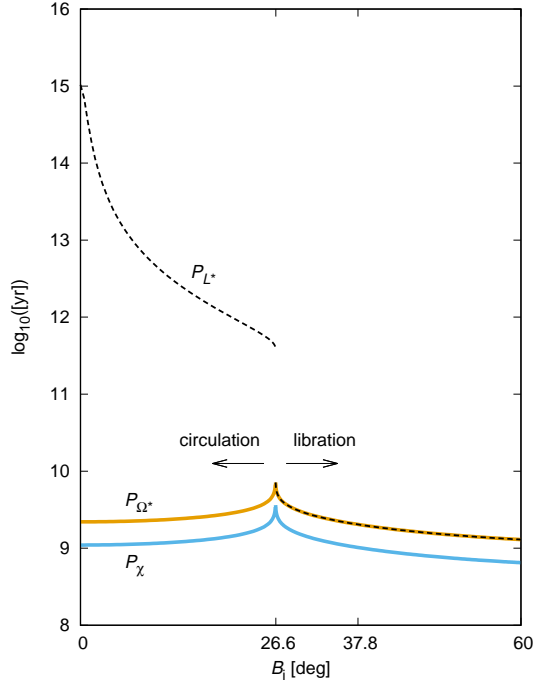}}
    \caption{
      Periods of $\chi$ (light blue), $\Omega^*$ (orange), and $L^*$ (black)
      given by Equations (\ref{eq:pchi}), (\ref{eq:plos}), and (\ref{eq:pls}),
      respectively,
      for $(a, q_i, i_i)=(2\times 10^4$ au, 10 au, 60$^\circ$)
      as a function of $B_i$.
    }
    \label{fig:bp}
  \end{center}
\end{figure}

From Equation (\ref{eq:chi}), we learn that
$\chi$ oscillates between the maximum ($\alpha_1$) and minimum ($\alpha_0$)
values according to the parameter $u$, with the period of $u=2K$, where $K=K(k)$ is
a complete elliptic integral of the first kind with modulus $k$.
The period given in time is
\begin{eqnarray}
  P_\chi = 2K\frac{g}{A_1}.
  \label{eq:pchi}
\end{eqnarray}
Figure \ref{fig:bp} shows periods of evolution against
$B_i$, the value of $B$ at $t=0$.
The light blue curve in Figure \ref{fig:bp} is
$P_\chi$ for $a=2\times10^4$ au, $q_i=10$ au, and $i_i=60^\circ$
obtained from Equation (\ref{eq:pchi}),
where $q_i$ and $i_i$ are the values of $q$ and $i$ at $t=0$, respectively.
It varies with $B_i$ and becomes a maximum at the separatrix.
However, the dependence on $B_i$ except around the separatrix
is not as strong as that on $a$,
which is proportional to $a^{-3/2}$.
The typical value of $P_\chi$ for bodies whose eccentricity at $t=0$ is $e_i\simeq 1$
is derived in Section \ref{ss:app}
as a function of $a$ (eq.(\ref{eq:pchi3})). 
The period of the oscillation of $i$ and that of $\omega$
in the case of libration are the same as $P_\chi$.
For $\omega$ in the case of circulation, the period of circulation is 2$P_\chi$.
The period of oscillation of $B$ is the same as that for $\omega$,
i.e., $P_\chi$ or 2$P_\chi$.

Using Equation (\ref{eq:chi}), one can calculate $e$ at time $t$
from $e_i$, $i_i$, and $\omega_i$.
Once $e$ is calculated, $i$ is obtained from Equation (\ref{eq:j})
and then $\sin^2\omega$ and $\sin B$ are obtained from
equations (\ref{eq:sin2o}) and (\ref{eq:sinb}), respectively.
To calculate the real $\omega$ from $\sin^2\omega$,
one needs to know if $\omega$ circulates or librates
from Equation (\ref{eq:kozai}) and at which phase of the evolution 
the time $t$ is located using the period given by Equation (\ref{eq:pchi}).
One can also use Equation (\ref{eq:cb}) instead of (\ref{eq:c}) to obtain $|B|$.

\subsubsection{Solution to $\Omega$}
Substituting Equations (\ref{eq:c}) and (\ref{eq:j}) into Equation (\ref{eq:dlodt})
and replacing $1-e^2$ with $\chi$, we have
\begin{equation}
  \frac{d\Omega}{dt}
  =-\frac{A_2}{1 - w_1^2{\rm sn}^2(u,k)},
  \label{eq:dlodt2}
\end{equation}
where
\begin{equation}
  A_2
  =\frac{A_1}{4}j\frac{\chi_0^*-j^2}{\alpha_0-j^2},
  \label{eq:A_2}
\end{equation}
\begin{equation}
  w_1^2=\frac{\alpha_1-\alpha_0}{j^2-\alpha_0}.
  \label{eq:w12}
\end{equation}
The integration of (\ref{eq:dlodt2}) with $t$ is rewritten as
\begin{eqnarray}
  \Omega
  &=&\Omega_i-A_2\int^{t'}_0\frac{dt}{1 - w_1^2{\rm sn}^2(u,k)}
  \nonumber\\
  &=&\Omega_i-A_3
  \int^{u'}_{u_0}\frac{du}{1 - w_1^2{\rm sn}^2(u,k)},
  \label{eq:lomega}
\end{eqnarray}
where $\Omega_i$ is the value of $\Omega$ at $t=0$ and
\begin{equation}
  u=\frac{A_1}{g}t+u_0, \quad dt= \frac{g}{A_1}du,
  \label{eq:dt}
\end{equation}
\begin{equation}
  A_3 = g\frac{A_2}{A_1}
  = \frac{1}{2}\frac{j(j^2-\chi_0^*)}{j^2-\alpha_0}\frac{1}{\sqrt{\alpha_2-\alpha_0}}.
  \label{eq:A_3}
\end{equation}
Since sn$^2(u,k)$ oscillates with a period of $u=2K$,
we split $u'$ as 
\begin{equation}
  u'=  u_r + 2mK,
  \label{eq:ud}
\end{equation}
where $m$ is an integer that gives $0\le u_r<2K$.
Then, Equation (\ref{eq:lomega}) is written and integrated
using an elliptic integral of the third kind $\Pi$ \citep{yellow} as
\begin{eqnarray}
  \Omega
  &=&\Omega_i-A_3
  \left\{
  \int^{u_r}_0\frac{du}{1 - w_1^2{\rm sn}^2(u,k)}
  +2m\int^{K}_0\frac{du}{1 - w_1^2{\rm sn}^2(u,k)}
  -\int^{u_0}_0\frac{du}{1 - w_1^2{\rm sn}^2(u,k)}
  \right\}
  \nonumber\\
  &=&\Omega_i-A_3
  \left[
    \Pi_\Omega'+ 2m\Pi(K,w_1^2,k)-\Pi(u_0,w_1^2,k)
    \right],
  \label{eq:lomega2}
\end{eqnarray}
where $\Pi(K, w_1^2, k)$ is a complete elliptic integral of the third kind and
\begin{eqnarray}
  \Pi_\Omega'=
  \left\{
  \begin{array}{lll}
    \Pi(u_r,w_1^2,k)
    &\quad{\rm for}&\;\;u_r<K\\
    2\Pi(K,w_1^2,k)-\Pi(2K-u_r,w_1^2,k)&\quad{\rm for}&\;\;u_r>K.
  \end{array}
  \right.
\end{eqnarray}

The period of $\Omega$
can be estimated by the linear approximation as
\begin{eqnarray}
  \Omega^*
  &=&\Omega_i -\frac{A_3\Pi[K,w_1^2,k]}{K} (u-u_0) 
  \nonumber\\
  &=&\Omega_i -n_{\Omega^*}t,
\end{eqnarray}
where
\begin{eqnarray}
  n_{\Omega^*}&=&\frac{A_2}{K}\Pi[K,w_1^2,k].
\end{eqnarray}
The period of $\Omega^*$ is obtained as
\begin{eqnarray}
  P_{\Omega^*}
  = \frac{2\pi K}{A_2\Pi[K,w_1^2,k]}.
  \label{eq:plos}
\end{eqnarray}
Figure \ref{fig:bp} shows $P_{\Omega^*}$ in orange against $B_i$
for $a=2\times10^4$ au, $q_i=10$ au, and $i_i=60^\circ$.
The behavior of $P_{\Omega^*}$ is quite similar to that of $P_\chi$.
The approximate relation between $P_{\Omega^*}$ and $P_\chi$
is shown in Section \ref{ss:app}.

\subsubsection{Solution to $L$}
From Equations (\ref{eq:sinb}), (\ref{eq:j}), and (\ref{eq:sin2o}),
$\sin^2 B$ is expressed with $c$, $j$, and $\chi$ as
\begin{equation}
  \sin^2 B\;=\;\sin^2\omega\sin^2 i = \frac{c+j^2-\chi}{5(1-\chi)}.
  \label{eq:sin2b}
\end{equation}
Then, $\tan^2 B$ is written as
\begin{equation}
  \tan^2 B=
  \frac{\sin^2 B}{1-\sin^2 B}
  =
  \frac{1}{4}\left[ 1+
  \frac{5(c+j^2-1)}{5-c-j^2-4\chi}
  \right].
  \label{eq:tan2b}
\end{equation}
Substituting Equation (\ref{eq:chi}) into Equation (\ref{eq:tan2b}),
$\tan^2 B$ is expressed as
\begin{equation}
  \tan^2 B
  =
  \frac{1}{4}\left[ 1+
    \frac{A_4}{1-w_2^2{\rm sn}^2(u,k)}
  \right],
  \label{eq:tan2b2}
\end{equation}
where
\begin{equation}
    A_4=\frac{5(c+j^2-1)}{5-c-j^2-4\alpha_0},
    \label{eq:A_4}
\end{equation}
\begin{equation}
  w_2^2=\frac{4(\alpha_1-\alpha_0)}{5-c-j^2-4\alpha_0}.
  \label{eq:w22}
\end{equation}
Substituting Equation (\ref{eq:tan2b}) into Equation (\ref{eq:dldt}),
we have
\begin{equation}  
  \frac{d L}{dt}
  =
  -\frac{A_1}{4}j-\frac{A_5}{1-w_2^2{\rm sn}^2(u,k)},
  \label{eq:dldt3}
\end{equation}
where
\begin{equation}
  A_5
  =\frac{1}{4}A_1A_4j
  =A_1\frac{5j(c+j^2-1)}{4(5-c-j^2-4\alpha_0)}.
  \label{eq:A_5}
\end{equation}  
As well as $\Omega$, 
the integration of (\ref{eq:dldt3}) with $t$ is expressed
using an elliptic integral of the third kind as
\begin{equation}
  L
  =
  L_i-A_7u
  -A_6\left[\Pi_{L}'+2m\Pi(K, w_2^2, k)
    -\Pi(u_0,w_2^2,k)
    \right],
  \label{eq:elu2}
\end{equation}
where $L_i$ is the value of $L$ at $t=0$ and
\begin{equation}
  A_6=g\frac{A_5}{A_1}
  =
  \frac{5j(c+j^2-1)}{2\sqrt{\alpha_2-\alpha_0}(5-c-j^2-4\alpha_0)},
  \label{eq:A_6}
\end{equation}
\begin{equation}
  A_7=\frac{j}{4}g=\frac{j}{2\sqrt{\alpha_2-\alpha_0}},
  \label{eq:A_7}
\end{equation}
\begin{eqnarray}
  \Pi_L'=
  \left\{
  \begin{array}{cll}
    \Pi(u_r,w_2^2,k)&\quad{\rm for}&\;\;u_r<K\\
    2\Pi(K,w_2^2,k)-\Pi(2K-u_r,w_2^2,k)&\quad{\rm for}&\;\;u_r>K.
  \end{array}
  \right.
\end{eqnarray}

The period of $L$ is also estimated in the same manner as $\Omega$
using the linear approximation, 
\begin{eqnarray}
  L^*
  &=&
  L_i-\frac{A_1}{4}jt
  -\frac{A_6\Pi[K,w_2^2,k]}{K} (u-u_0),
  \nonumber\\
  &=&
  L_i-n_{L^*}t,
\end{eqnarray}
where
\begin{equation}
  n_{L^*}
  =\frac{A_1}{4}j\left(1+\frac{A_4}{K}\Pi[K,w_2^2,k]\right).
\end{equation}
The period of $L^*$ is obtained as
\begin{eqnarray}
  P_{L^*}
  =  \frac{8\pi}{A_1j}\left(1+\frac{A_4}{K}\Pi[K,w_1^2,k]\right)^{-1}
  \label{eq:pls}
\end{eqnarray}
The black dashed curve in Figure \ref{fig:bp} shows $P_{L^*}$
plotted against $B_i$
for $a=2\times10^4$ au, $q_i=10$ au, and $i_i=60^\circ$.
The behavior of $P_{L^*}$ looks identical to that of $P_{\Omega^*}$ for $B_i>26^\circ.6$.
In contrast, for $B_i<26^\circ.6$, $P_{L^*}$ suddenly becomes
$10^{2-5}$ times larger than that for $B_i>26^\circ.6$.
For this example, $P_{L^*}$ for $c<1$ is much longer than
the age of the solar system.

%\newpage
\section{Evaluation of analytical solutions with numerical calculations}
\label{ss:num}
We test the analytical solutions
to the orbital elements, $L$, and $B$ by comparing with
the orbital evolution obtained by numerical integrations.
We are especially interested in checking two approximations that
we have made in the derivation of the analytical solutions:
the axisymmetric approximation of the Galactic tide
by neglecting the radial component and
the time-averaging of the Hamiltonian
assuming that the Galactic tide is much smaller than
the solar gravity.
\citet{2001MNRAS.324.1109B} examined both approximations
using numerical orbital integrations
for comets mainly with $e\ll 1$.
\citet{2007AJ....134.1693H} considered
comets with $e_i\sim 1$ and showed that
the time-averaging is plausible for comets with orbital periods
$T_{\rm K}\lesssim 10P_\chi$;
however, they neglected the radial component of the Galactic tide
in their numerical integrations.
In this paper, we focus on comets with $e\simeq 1$ and
examine how the analytical solutions are useful
for the discussion about long-period comets in the observable region.

\subsection{Equation of Motion}
Under the epicyclic approximation \citep[e.g.,][]{1987gady.book.....B},
the equation of motion of a body orbiting around the Sun with
tidal forces from the Galactic disk is
\begin{equation}
  \frac{d{\bf r}}{dt} = -GM_\odot\frac{\bf r}{r^3}+f_{\rm tide},
  \label{eq:em}
\end{equation}
where ${\bf r}$ is the position of the body with respect to the Sun
and $f_{\rm tide}$ is the Galactic tide,
\begin{equation}
  f_{\rm tide} = \Omega_0^2(x'-y') -\nu_0^2z',
  \label{eq:ftide}
\end{equation}
where $x'$, $y'$, and $z'$ give the position of the body
in rotating coordinates centered on the Sun, $\Omega_0$
is the circular frequency (i.e., the angular speed of the Sun in the Galaxy),
$\nu_0=\sqrt{4\pi G\rho}$ is the vertical frequency,
and $\rho$ is the total density in the solar neighborhood.
We adopt $\Omega_0=26$ km s$^{-1}$ \citep[e.g.,][]{1987gady.book.....B}
and $\rho=0.1\;M_\odot{\rm pc}^{-3}$ \citep{2000MNRAS.313..209H},
which give $\nu_0^2/\Omega_0^2\simeq 10$.
To evaluate the analytical solutions derived in Section \ref{ss:ana},
we integrated Equation (\ref{eq:em}) for bodies initially on the ecliptic plane
(i.e., $i_i=60^\circ$, $\Omega_i=186^\circ$)
for 4.5 Gyr with the $P(EC)^2$ Hermite scheme \citep{1998MNRAS.297.1067K}
and compared the orbital evolutions with the analytical solutions.
The bodies are set at their perihelion at $t=0$.
We also performed extra numerical calculations that neglect the
first term in Equation (\ref{eq:ftide})
to examine the effect of the radial component of the Galactic tide.

\subsection{Comparison}
In this section, we compare the results of numerical calculations
that consider both the radial and vertical components of the Galactic tide
and the analytical solutions by plotting them together
in the same figures.

\begin{figure}[hbtp]
  \begin{center}
   \resizebox{15cm}{!}{\includegraphics{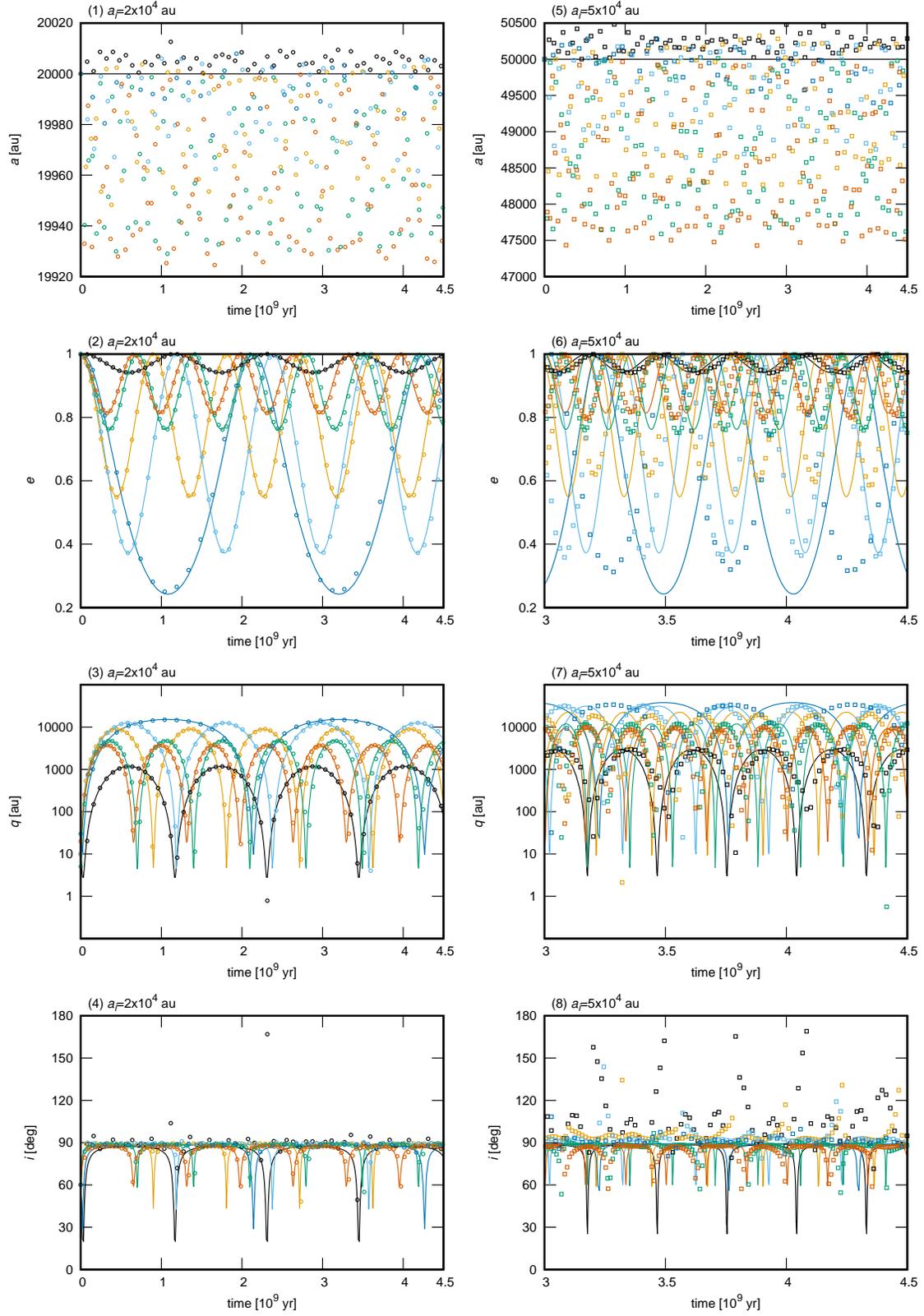}}
    \caption{
      Evolution of $a$, $e$, $q$, and $i$ of
      bodies orbiting around the Sun with
      the tidal forces from the Galactic disk.
      Circles/squares are obtained by numerical integration of
      Equation (\ref{eq:em}) and the solid curves are analytical solutions.
      Left and right panels are for bodies with $a_i=2\times10^4$ au
      (circles) and $a_i=5\times10^4$ au (squares), respectively.
      All bodies have
      $(i_i, \Omega_i)=(60^\circ, 186^\circ)$.
      For $q_i$ and $\omega_i$,
      $(q_i, \omega_i)=$ 
      (black: 10 au, 10$^\circ$),
      (orange: 10 au, 130$^\circ$),
      (light blue: 10 au, 320$^\circ$),
      (green: 5 au, 70$^\circ$),
      (dark orange: 20 au, 280$^\circ$), and
      (blue: 30 au, 210$^\circ$).
      The range of the $x$ axis for bodies with
      $a_i=5\times10^4$ au (right panels)
      is from 3 to 4.5 Gyr except in panel (5).
    }
    \label{fig:t_oe1}
  \end{center}
\end{figure}

\begin{figure}[hbtp]
  \begin{center}
   \resizebox{15cm}{!}{\includegraphics{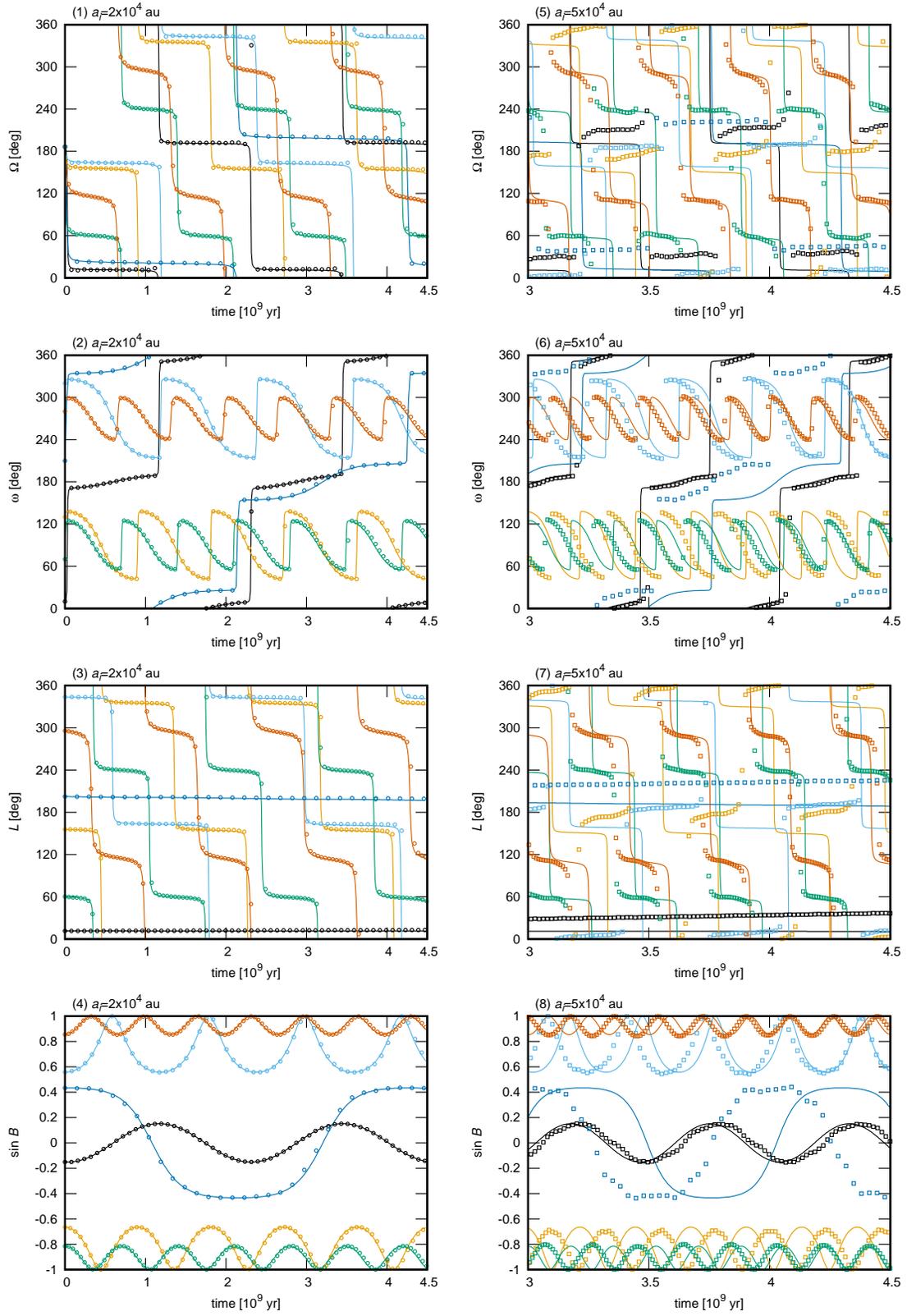}}
    \caption{
      Evolution of $\Omega$, $\omega$, $L$, and $\sin B$ of
      the same bodies shown in Figure \ref{fig:t_oe1}.
      The range of the horizontal axis for bodies with
      $a_i=5\times10^4$ au (right panels)
      is from 3 to 4.5 Gyr.
    }
    \label{fig:t_oe2}
  \end{center}
\end{figure}

Figures \ref{fig:t_oe1} and \ref{fig:t_oe2}
show the orbital evolutions of bodies against time for 4.5 Gyr.
All bodies have
$(i_i, \Omega_i)=(60^\circ, 186^\circ)$
but different values (color-coded) of $q_i$ and $\omega_i$.
Bodies in the same colors on the left and right panels
have the same initial orbital elements except for $a_i$, 
the semimajor axis at $t=0$;
$a_i=2\times 10^4$ au and $5\times 10^4$
for the left and right panels, respectively.
Circles and squares are the results obtained by numerical integration of Equation (\ref{eq:em})
and the solid curves are the analytical solutions derived in Section \ref{ss:ana}.
Panel (1) in Figure \ref{fig:t_oe1} shows the evolution of $a$.
There are variations within $\lesssim 0.5\%$,
but no systematic decay or increase is seen for 4.5 Gyr evolution.
We find the same features in results of numerical calculations
without the radial component of the Galactic tide.
We conclude that
the variation of $a$ is due to the short-term effect
of the Galactic tide that is neglected in the time-averaging process
in the derivation of Equation (\ref{eq:h}).

Panel (2) in Figure \ref{fig:t_oe1} shows the evolution of $e$.
Oscillations with various periods and amplitudes occur
since the Hamiltonian of a body depends on $\omega_i$
as seen in Equation (\ref{eq:c}).
The initial values are very close to 1 because
$e_i=1-q_i/a_i=0.9985\;-\;0.99975$,
but can be larger if $\sin^2\omega_i\ne 1$.
Panel (3) in Figure \ref{fig:t_oe1} shows the evolution of $q$.
The range of the time when the bodies arrive in the observable region,
i.e., $q\lesssim 10^2$ au, is very short compared to the period
of the oscillation.
For most of the time, $q$ is outside the planetary region and planetary
perturbations are negligible for those orbits.
The analytical solutions and the results of numerical calculations
shown in panels (2) and (3) are in good agreement.
In contrast, those for the evolution of $i$
are not in good agreement as shown in panel (4) in Figure \ref{fig:t_oe1}.
For the analytical solutions, the orbits never become retrograde
as a consequence of the conservation of $j$.
Panels (5)-(8) in Figure \ref{fig:t_oe1} are the same as
panels (1)-(4) but for $a_i=5\times 10^4$ au.
Note that only for $t>3$ Gyr are results shown, except in panel (5).
They have the same features as for $a_i=2\times 10^4$ au,
although the agreement of the analytical solutions
with the results of the numerical calculations
is worse because of the shift of
the oscillation phase.
The shift can simply be explained by the evolution of $a$,
which is not completely equal to $a_i$ as shown in panel (1) in
Figure \ref{fig:t_oe1}.
This makes the period of oscillation slightly
longer/shorter than the period given by Equation (\ref{eq:pchi}).
Since the variation of $a$ is larger for large $a_i$
and the dynamical evolution is quicker for large $a_i$,
the shift of periods is not negligible in 4.5 Gyr for $a_i=5\times 10^4$ au.
In contrast, the amplitudes of the oscillations of $e$ and $q$
found in the numerical calculations show
rather good agreement with the analytical solutions.

\begin{figure}[hbtp]
  \begin{center}
    \resizebox{13cm}{!}{\includegraphics{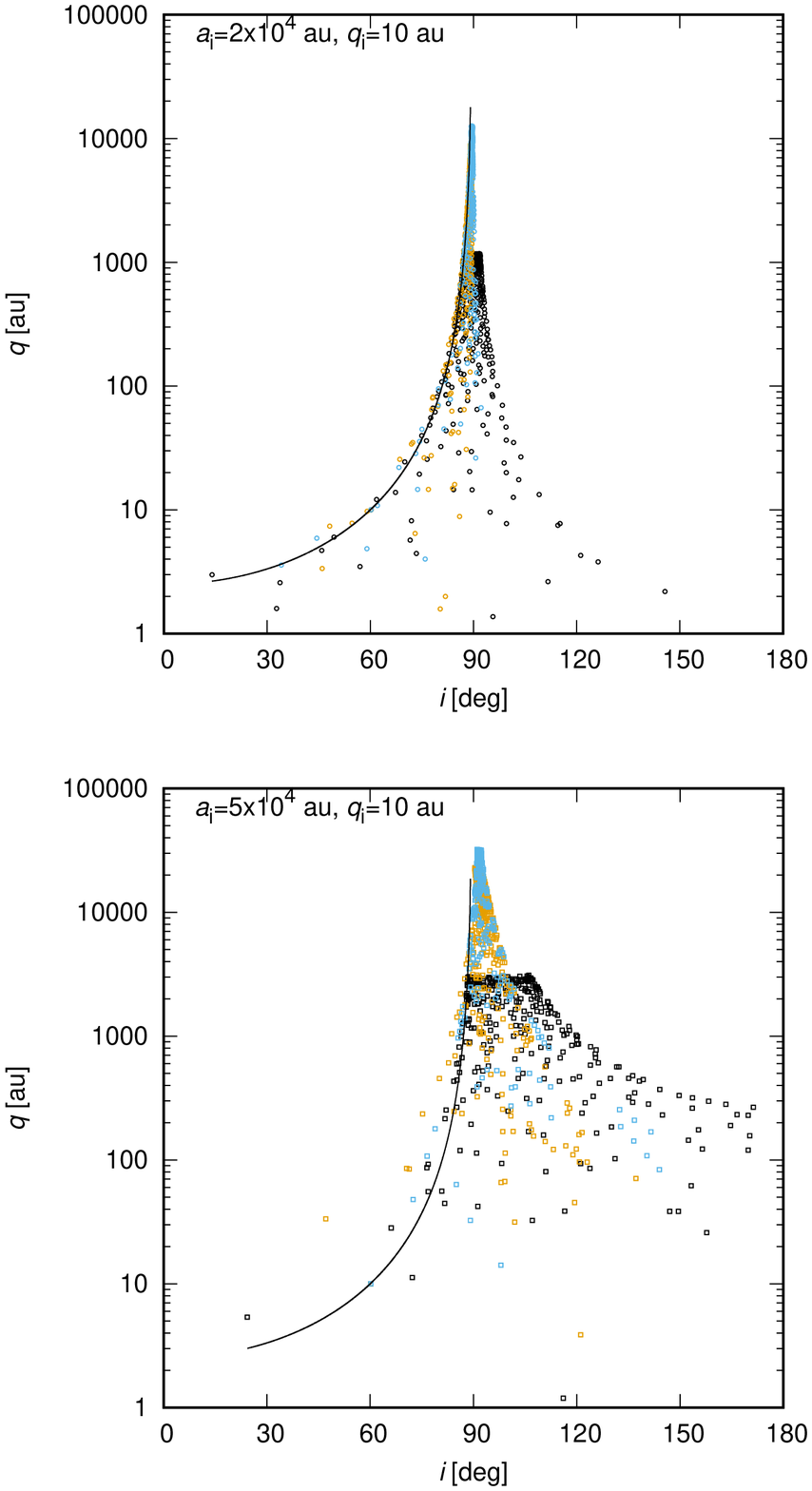}}
    \caption{
      Orbital evolution on the $i-q$ plane.
      Circles show the results of numerical calculation
      of the equation of motion given by Equation (\ref{eq:em})
      for 4.5 Gyr
      for bodies with
      $(a_i, q_i, i_i)=(2\times 10^4$ au, 10 au, 60$^\circ$) (top),
      $(a_i, q_i, i_i)=(5\times 10^4$ au, 10 au, 60$^\circ$) (bottom),
      and with $\omega_i=$ 10$^\circ$ (black), 130$^\circ$ (orange),
      and 320$^\circ$ (light blue). 
      A solid curve in each panel shows
      the equi-$j$ curve that all the points would be on
      if $j$ were completely conserved.
    }
    \label{fig:iq}
  \end{center}
\end{figure}

The top panel in Figure \ref{fig:iq} shows orbital evolution
on the $i-q$ plane with the same symbols and colors
as in panels (1)-(4) in Figure \ref{fig:t_oe1}.
All the bodies initially have the value of $j$ that is shown
as an equi-$j$ curve (black solid) given by Equation (\ref{eq:j}),
where $e=1-q/a$ is substituted.
The results of numerical calculations shown with circles
are scattered away from the equi-$j$ curve for small $q$.
Consequently, $j$ is not conserved completely.
However, for very small $q$ (i.e., $e\sim 1$),
$j$ is very small independently of $i$.
In other words, $j$ is approximately conserved for 4.5 Gyr.
Consequently, $i$ always reaches $\simeq 90^\circ$ when $q$ is large.
The bottom panel shows the same as the top one but for $a_i=5\times 10^4$ au.
The conservation of $j$ is worse than for $a_i=2\times 10^4$ au
but still we can say it is approximately conserved even when
the radial component of the Galactic tide is included.

Panel (1) in Figure \ref{fig:t_oe2} shows the evolution of $\Omega$.
The evolution is characterized as a decreasing step function.
The values at each step are different among the bodies
although they all have the same value of $\Omega_i=186^\circ$.
As is clear from Equation (\ref{eq:dlodt}),
the big drop in $\Omega$ occurs when $e$ is large, i.e., $q$ is small.
Panel (2) in Figure \ref{fig:t_oe2} shows the evolution of $\omega$.
Two of the six bodies, those with $\omega_i=10^\circ$ (black) and 210$^\circ$ (blue),
are in the case of circulation.
Their behavior of having a big change at $e\simeq 1$
is quite similar to that of $\Omega$ but they increase with time.
The other four bodies are in the case of libration
and they librate around $\omega=90^\circ/270^\circ$ depending on each $\omega_i$.
Panel (3) in Figure \ref{fig:t_oe2} shows the evolution of $L$,
which is expressed as a function of $i$, $\Omega$, and $\omega$
(eq. (\ref{eq:elu})).
Interestingly, for bodies in the case of circulation,
$L$ is almost constant beyond 4.5 Gyr.
For bodies in the case of libration, their evolutions are quite
similar to those of $\Omega$.
However, the phases are shifted so that the value of $L$
is almost constant when $q\lesssim 10^2$ au.
Panel (4) in Figure \ref{fig:t_oe2} shows the evolution of $\sin B$,
which is expressed as a function of $i$ and $\omega$ (eq.(\ref{eq:sinb})).
For bodies in the case of circulation, $\sin B$ oscillates
symmetrically with respect to $\sin B=0$, the Galactic plane.
As given by Equation (\ref{eq:conb}), the cases of circulation and
libration are divided by $\sin |B_i|=\sqrt{1/5}=0.447$.
In panels (1)-(4),
the analytical solutions and the results of numerical calculations
are in good agreement.
Panels (5)-(8) in Figure \ref{fig:t_oe2} are the same as
panels (1)-(4) but for $a_i=5\times 10^4$ au.
Just as seen in Figure \ref{fig:t_oe1},
the agreement of the analytical solutions with
the results of numerical calculations is worse.
In particular, in panels (5) and (7),
four of six bodies show an increase in $\Omega$ and $L$
in the numerical calculations,
which is never given by the analytical solutions.
From several extra numerical calculations, we confirmed that these opposite
evolutions seen in $\Omega$ and $L$ are due to the radial component
of the Galactic tide that breaks the conservation of $j$.
In panels (6) and (8),
the disagreement that arises from the shifts of the periods is quite significant;
however, the amplitudes of the oscillations show good agreement even
after 4.5 Gyr.

From the above comparisons, we conclude that
the analytical solutions are basically useful for describing the
orbital evolution, except for $i$ and $\Omega$ of comets
in the observable region (i.e., $q\lesssim 10^2$ au).
For bodies with $a=5\times 10^4$ au, the small differences
between the periods given by the analytical solutions
and the ones obtained from the numerical integrations
pile up and are quite large at $t\sim$ a few Gyr.
This could be understood simply as the results of the shifts
of oscillation/circulation of orbital evolution.
Therefore, the time evolution normalized by the periods
obtained by numerical integrations is well reproduced
by the analytical solutions.

%\newpage
\section{Quasi-Rectilinear approximation and application to long-period comets}
\label{ss:app}
In this section, we apply the analytical solutions
derived in Section \ref{ss:ana}
to fictional observable long-period comets entering
the planetary region from the Oort cloud.
We assume that the comets initially
have very elongated orbits given by planetary scattering
on the ecliptic plane.
For these comets, setting $e_i\simeq 1$ is a good approximation
and it makes the solutions simple.
We call this approximation as the quasi-rectilinear approximation.
Since the planetary scattering does not give high inclinations
(see Appendix),
we assume $i_i= i_\odot=60^\circ$
and $\Omega_i=\Omega_\odot=186^\circ$
to be on the ecliptic plane.
Using this approximation, we compare the periods of
$\chi$, $\Omega$, and $L$ and investigate
the relation among them especially for comets in the observable region.

\subsection{Preparation}
In this section, we first derive the explicit expression for
$\alpha_0$, $\alpha_1$, and $\alpha_2$ (eq.(\ref{eq:aaa}))
and their relation, $k^2$, $w_1^2$, and $w_2^2$
(eqs. (\ref{eq:k2}), (\ref{eq:w12}), and (\ref{eq:w22}))
under the quasi-rectilinear approximation, which gives $0<j^2\ll 1$.
Using the expressions,
we calculate the values of $\Pi$ that
appear in solutions to $\Omega$ and $L$.

\subsubsection{solutions and parameters}
Substituting $e=e_i$, $B=B_i$, and $1-e_i^2=j^2/\cos^2 i_\odot=4j^2$
into Equation (\ref{eq:cb}),
we obtain 
\begin{equation}
  c = 3j^2+5(1-4j^2)\sin^2 B_i.
  \label{eq:cb2}
\end{equation}
Using $-i_\odot\le B_i\le i_\odot$,
we have the minimum and maximum values of $c$ as
\begin{eqnarray}
  \begin{array}{clll}
  c_{\rm min}&=& 3j^2 \simeq 0&\;\;\mbox{\rm for}\;\; B_i=0
  \\
  c_{\rm max}&=& 3j^2+\frac{15}{4}(1-4j^2)
  \simeq \frac{15}{4}&\;\;\mbox{\rm for}\;\; B_i=i_\odot
  \end{array}.
  \label{eq:cmnmx}
\end{eqnarray}
where $j^2\ll 1$ is used.

From Equation (\ref{eq:qq}), the explicit expressions of the solutions are
approximated as
\begin{eqnarray}
  \chi_1^*
  &\simeq&
  \frac{5j^2}{5+4j^2-c}
  +\frac{100j^4}{\left(5+4j^2-c\right)^3}
  +{\mathcal O}(j^6)
  \ll 1
  \label{eq:chi1}
  \\
  \chi_2^*
  &\simeq&
  \frac{5+4j^2-c}{4}-\frac{5j^2}{5+4j^2-c}  +{\mathcal O}(j^4)
  \label{eq:chi2}
\end{eqnarray}
where $j^2\ll 1$ is used but the term ${\mathcal O}(j^4)$ in Equation (\ref{eq:chi1})
is left for the comparison with $j^2$.

To evaluate the relation between $j^2$ and $\chi_1^*$, we substitute $c=c_{\rm min}$
since the difference between $j^2$ and $\chi_1^*$ becomes minimum for $c=c_{\rm min}$.
It is calculated as
\begin{eqnarray}
  \chi_{1}^* -j^2
  &\simeq&
  \frac{5j^2}{5+4j^2-c_{\rm min}}
  +\frac{100j^4}{\left(5+4j^2-c_{\rm min}\right)^3}
  -j^2
  \nonumber\\
  &=&
  \frac{j^4(75-10j^2-j^4)}{(5+j^2)^3}\;>\;0.
  \label{eq:c1-j2}
\end{eqnarray}
Therefore, the relation between $j^2$ and $\chi_{1}^*$ is always as $j^2<\chi_1^*$.

From Equations (\ref{eq:chi0}), (\ref{eq:cmnmx}), and (\ref{eq:chi1}), 
the relation between $\chi_0^*$ and $\chi_{1}^*$ is always as $\chi_1^*<\chi_0^*$.
The relation between $\chi_0^*$ and $\chi_2^*$ depends on the value of $c$.
The difference is calculated as
\begin{eqnarray}
  \chi_0^*-\chi_2^*\simeq c -\frac{5-c}{4}
  =\frac{5(c-1)}{4},
  \label{eq:c0-c2}
\end{eqnarray}
where the terms ${\mathcal O}(j^2)$ are neglected.
As the value of $c$ at the separatrix (eq. (\ref{eq:kozai}))
is approximated as $c+j^2\simeq c$,
the relation is approximately given as
\begin{eqnarray}
  \begin{array}{lll}
    \chi_0^*<\chi_2^* &\;\;{\rm for}\;\;& c<1 \;\;{\rm (circulation)}\\
    \chi_0^*>\chi_2^* &\;\;{\rm for}\;\;& c>1 \;\;{\rm (libration)}
  \end{array}.
  \label{eq:c0c2}
\end{eqnarray}

Summarizing the relations, we have
\begin{eqnarray}
  (\alpha_0,\alpha_1,\alpha_2,)=
  \left\{
  \begin{array}{lll}
    (\chi_1^*, \chi_0^*, \chi_2^*)
    &\;\;{\rm for}\;\;& c<1 \;\;{\rm (circulation)}\\
    (\chi_1^*, \chi_2^*, \chi_0^*)
    &\;\;{\rm for}\;\;& c>1 \;\;{\rm (libration)}
  \end{array}
  \right.
  \label{eq:aaa}
\end{eqnarray}
and $\alpha_1=\alpha_2$ for the separatrix for $c=1$.

Using Equations (\ref{eq:k2}) and (\ref{eq:aaa}) and $j^2\ll 1$, we have
\begin{eqnarray}
  k^2=
  \left\{
  \begin{array}{cll}
    \frac{4c}{5-c}\;<\;1
    &\;\;{\rm for}\;\;& c<1 \;\;{\rm (circulation)}\\
    \frac{5-c}{4c}\;<\;1
    &\;\;{\rm for}\;\;& c>1 \;\;{\rm (libration)}
  \end{array}
  \right.
  \label{eq:kkk}
\end{eqnarray}
and $k^2=1$ for the separatrix for $c=1$.

Using $j^2<\alpha_0\ll 1$ and Equations (\ref{eq:c1-j2}) and (\ref{eq:aaa}),
the parameter $w_1^2$ that appears in $\Pi$
for $\Omega$ given by Equation (\ref{eq:w12}) is expressed as
\begin{equation}
  w_1^2=\frac{\alpha_1-\alpha_0}{j^2-\alpha_0}\simeq -\infty,
  \label{eq:w12_2}
\end{equation}
and another parameter $w_2^2$, which is for $L$
given by Equation (\ref{eq:w22}), is 
\begin{eqnarray}
  w_2^2\simeq
  \left\{
  \begin{array}{cll}
    \frac{4c}{5-c}
    &\;\;{\rm for}\;\;& c<1 \;\;{\rm (circulation)}\\
    1
    &\;\;{\rm for}\;\;& c>1 \;\;{\rm (libration)}
  \end{array}.
  \right.
  \label{eq:w22_2}
\end{eqnarray}

\subsubsection{Complete integrals of the third kind in $\Omega$ and $L$}
Depending on the values and relation between $k^2$ and $w_1^2$ or $w_2^2$,
the complete elliptic integrals of the third kind are expressed in
different forms \citep{yellow}.

For $0<-w_1^2<\infty$, which is called "case I" in \citet{yellow},
\begin{equation}
  \Pi(w_1^2,k)
  =\frac{k^2K}{k^2-w_1^2}-\frac{\pi w_1^2\Lambda_0(\varphi,k)}{2\sqrt{w_1^2(1-w_1^2)(w_1^2-k^2)}}
  \quad (w_1^2<0),
  \label{eq:ce3}
\end{equation}
where
\begin{equation}
  \Lambda_0(\varphi,k)=\frac{2}{\pi}[EF(\varphi,k')+KE(\varphi,k')-KF(\varphi,k')],
  \label{eq:lam}
\end{equation}
where $E=E(k)$ and $E(\varphi, k')$ are a complete and normal elliptic integrals
of the second kind,
\begin{equation}
  \sin\varphi=\sqrt{\frac{w_1^2}{w_1^2-k^2}},
  \label{eq:sinvp}
\end{equation}
and
\begin{equation}
  k'^2=1-k^2.
  \label{eq:kd2}
\end{equation}
Using Equation (\ref{eq:w12_2}), we have $\sin\varphi=1$.
Then $F(\varphi, k')$ and $E(\varphi, k')$ in Equation (\ref{eq:lam}) become 
complete elliptic integrals of the first and second kinds, $K'=K(k')$ and $E'=E(k')$,
respectively, and
\begin{equation}
  \Lambda_0=\frac{2}{\pi}[EK'+E'K-KK']=\frac{2}{\pi},
\end{equation}
which is called Legendre's relation.
Then $\Pi(w_1^2,k)$ is approximated as
\begin{equation}
  \Pi(w_1^2,k)
  \simeq \frac{\pi}{2}\frac{1}{\sqrt{-w_1^2}}.
  \label{eq:pi1}
\end{equation}

For $k^2<w_2^2<1$, which is called "case II" in \citet{yellow},
\begin{equation}
  \Lambda_0(\vartheta,k)=\frac{2}{\pi}[EF(\vartheta,k')+KE(\vartheta,k')-KF(\vartheta,k')],
\end{equation}
\begin{equation}
  \sin\vartheta=\sqrt{\frac{1-w_2^2}{k'^2}}
\end{equation}  
and
\begin{equation}
  \Pi(w_2^2, k)
  =K+\frac{\pi w_2[1-\Lambda_0(\vartheta,k)]
  }{2\sqrt{(w_2^2-k^2)(1-w_2^2)}}.
\end{equation}
For the special case of $w_2^2=k^2$, which is true for the case for circulation,
\begin{eqnarray}
  \Pi(k^2, k)=\frac{E}{1-k^2}.
\end{eqnarray}
For $w_2^2\simeq 1$ in the case of circulation, we have $\sin\vartheta=0$
and then
\begin{equation}
  \Lambda_0(\vartheta,k)\simeq 0.
\end{equation}
Therefore, $\Pi(w_2^2, k)$ is approximated as
\begin{eqnarray}
  \Pi(w_2^2, k)
  &\simeq&
  \left\{
  \begin{array}{cll}
    \frac{E}{1-k^2}
    &\;\;{\rm for}\;\;& c<1 \;\;{\rm (circulation)}\\
    K+\frac{\pi w_2}{2\sqrt{(w_2^2-k^2)(1-w_2^2)}}
    &\;\;{\rm for}\;\;& c>1 \;\;{\rm (libration)}.
  \end{array}
  \right.
  \label{eq:pi2}
\end{eqnarray}

\subsection{Periods of $\chi$, $\Omega$, and $L$ and their ratios}
\subsubsection{mean $P_\chi$}
Assuming a uniform distribution of $\omega_i$ between 0$^\circ$ and 360$^\circ$,
the mean value of $\sin^2\omega_i$ is
$\langle\sin^2\omega_i\rangle=1/2$, which corresponds to
$B_i\simeq 37^\circ.8$.
Using $i_i=60^\circ$ and $\sin^2\omega_i=1/2$,
the mean value of $c$ given by Equation (\ref{eq:c}) is approximated as
\begin{equation}
  \langle c\rangle\sim \frac{15}{8}.
  \label{eq:mc}
\end{equation}
Since $\langle c\rangle>1$, it is in the case of libration of $\omega$.
Therefore, from Equations (\ref{eq:aaa}) and (\ref{eq:kkk}), we have 
\begin{equation}
  \alpha_2=\chi^*_0\simeq\langle c\rangle+j^2,\quad
  k^2=\frac{5-\langle c\rangle}{4\langle c\rangle}.
  \label{eq:mk2}
\end{equation}
Substituting Equations (\ref{eq:mc}) and (\ref{eq:mk2}) and
the expansion in series of $K$ given as 
\begin{equation}
  K=\frac{\pi}{2}\left[1+\frac{1}{4}k^2+\frac{9}{64}k^4\ldots\right] \quad(k^2<1)
  \label{eq:exK}
\end{equation}
into Equation (\ref{eq:pchi}) and using $j^2\ll 1$,
the mean value of $P_\chi$ for a given $a$ and $\rho_0$ is estimated as
\begin{eqnarray}
  \langle P_\chi\rangle \simeq \frac{4K}{A_1\sqrt{\alpha_2}}
  \sim 10^9
  \left(\frac{0.1}{\rho\;[M_\odot{\rm pc}^{-3}]}\right)
  \left(\frac{2\times10^4}{a\;[{\rm au}]}\right)^{3/2}\;[{\rm yr}].
  \label{eq:pchi3}
\end{eqnarray}

\subsubsection{Ratio of $P_{\Omega^*}$ to $P_\chi$}
\begin{figure}[hbtp]
  \begin{center}
    \resizebox{8cm}{!}{\includegraphics{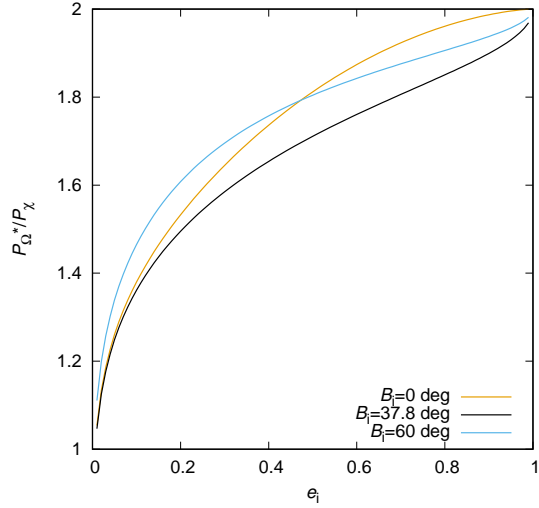}}
    \caption{
      Ratios of $P_{\Omega^*}$ to $P_\chi$ for $B=$0 (orange), 38$^\circ$ (black),
      and $60^\circ$ (light blue)
      and for $i_i=60^\circ$ as a function of $e_i$.
    }
    \label{fig:e-r}
  \end{center}
\end{figure}

Substituting Equations (\ref{eq:A_2}) and (\ref{eq:pi1}) into
equation (\ref{eq:plos}), the period of $\Omega$ is approximated as
\begin{equation}
  P_{\Omega^*}\simeq
  \frac{16K}{A_1}
  \frac{\sqrt{(\alpha_1-\alpha_0)(\alpha_0-j^2)}}{jc}.
  \label{eq:plos2}
\end{equation}
Using Equations (\ref{eq:g}), (\ref{eq:pchi}), and (\ref{eq:plos2}),
the ratio of $P_{\Omega^*}$ to $P_\chi$ is given as
\begin{equation}
  \frac{P_{\Omega^*}}{P_\chi}
  \simeq \frac{4\sqrt{\alpha_0\alpha_1\alpha_2-\alpha_1\alpha_2j^2}}{jc},
  \label{eq:ratio1}
\end{equation}
where $j^2<\alpha_0\ll 1$.
Using Equations (\ref{eq:chi0}), (\ref{eq:chi1}), (\ref{eq:chi2}), and (\ref{eq:aaa}),
the terms in the square root are calculated as
\begin{equation}
  \alpha_0\alpha_1\alpha_2=\chi_0^*\chi_1^*\chi_2^*
  \simeq
  \frac{5cj^2}{4}
  \label{eq:root1}
\end{equation}
and
\begin{equation}
  \alpha_1\alpha_2j^2
  = \chi_1^*\chi_2^*j^2
  \nonumber\\
  \simeq
  \frac{cj^2(5-c)}{4}.
  \label{eq:root2}
\end{equation}
Substituting Equations (\ref{eq:root1}) and (\ref{eq:root2}) into Equation (\ref{eq:ratio1}),
we have
\begin{equation}
  \frac{P_{\Omega^*}}{P_\chi}\simeq 2.
  \label{eq:ratio1_2}
\end{equation}
Consequently, not only $e$, $i$, and  $\omega$,
but also $\Omega$ displays a coupled evolution.
This commensurability is established only under the
quasi-rectilinear approximation.
Figure \ref{fig:e-r} shows the ratios of $P_{\Omega^*}$
to $P_\chi$ against $e$ for $B_i=0$, 37$^\circ$.8, and
60$^\circ$.
For any $B_i$, the ratio approaches 2 only when $e\sim 1$.
Note that \citet{2007AJ....134.1693H} already reported the commensurability
but they did not show it in equations.
Using the quasi-rectilinear approximation,
equation (\ref{eq:lomega}) is rewritten as
\begin{equation}
  \Omega
  \simeq\Omega_i-A_3\left[\Pi_\Omega'-\Pi(u_0,w_1^2,k)\right] -m\pi.
  \label{eq:lomega3}
\end{equation}

\subsubsection{Ratio of $P_{L^*}$ to $P_\chi$}
Substituting Equations (\ref{eq:A_4}) and (\ref{eq:pi2}) into
equation (\ref{eq:pls}), the period of $L$ is approximated as
\begin{eqnarray}
  P_{L^*}
  &\simeq&
  \left\{
  \begin{array}{cll}
    \frac{8\pi}{A_1j}\left[1+\frac{E}{K}\right]^{-1}
    &\;\;{\rm for}\;\;& c<1 \;\;{\rm (circulation)}\\
    \frac{8K}{A_1\sqrt{c}}
    &\;\;{\rm for}\;\;& c>1 \;\;{\rm (libration)}
  \end{array}
  \right.
  \label{eq:pls2}
\end{eqnarray}
Using Equations (\ref{eq:pchi}) and (\ref{eq:pls2}),
we have the ratios of $P_{L^*}$ to $P_\chi$ as
\begin{eqnarray}
  \frac{P_{L^*}}{P_\chi}
  &\simeq&
  \left\{
  \begin{array}{cll}
    \infty
    &\;\;{\rm for}\;\;& c<1 \;\;{\rm (circulation)}\\
    2
    &\;\;{\rm for}
    \;\;& c>1 \;\;{\rm (libration)}
  \end{array}
  \right.
  \label{eq:ratio2}
\end{eqnarray}
where $j^2\ll 1$, $0<E/K<1$, and $\alpha_2\simeq c$ for $c>1$ are used.
Then Equation (\ref{eq:elu2}) is rewritten as
\begin{eqnarray}
  L
  &\simeq&
  \left\{
  \begin{array}{cll}
  L_i
  &\;\;{\rm for}\;\;& c<1 \;\;{\rm (circulation)}\\
  L_i-A_6\left[\Pi_{L}'-\Pi(u_0,w_2^2,k)\right]-m\pi
  &\;\;{\rm for}\;\;& c>1 \;\;{\rm (libration)}
  \end{array}.
  \right.
  \label{eq:elu3}
\end{eqnarray}

The almost fixed value of $L$ beyond 4.5 Gyr
for the case of circulation already seen in Section \ref{ss:ana}
is also explained with Equations (\ref{eq:elu}), (\ref{eq:theta}),
and (\ref{eq:ratio1_2}) as follows.
Since the evolution of $i$ is a periodic oscillation,
the effect of $\cos i$ multiplied by $\tan\omega$
in Equation (\ref{eq:theta})
can be assumed null when it is averaged over the period,
i.e., as if $L\sim \Omega+\omega$.
Under the quasi-rectilinear approximation, 
$\Omega$ and $\omega$ for the case of circulation
evolve with the same velocities but in opposite directions.
Consequently, $\Omega$ and $\omega$ are canceled out and $L$ evolves little.

For the case of libration, the second term in Equation (\ref{eq:theta})
would be 0 on average over a period of libration.
As a result, $\Omega$ and $L$ evolve with the same period,
which is twice $P_\chi$.
This relation is true for any $e_i$ and does not require
the quasi-rectilinear approximation.

\subsection{Behavior in the observable region}
As confirmed in Section \ref{ss:num},
the time for $q\sim q_i$ is expressed as $t\simeq mP_\chi$,
where $m$ is an integer.
Using this relation and Equations (\ref{eq:chi}), (\ref{eq:lomega3}),
and (\ref{eq:elu3}),
we can use the combinations of orbital elements 
at $t=mP_\chi$.
as the prediction for $q\simeq q_i\lesssim 10^2$ au.
Table \ref{tb:1} summarizes the relation.
\begin{table}[hbtp]
  \begin{center}
    \begin{tabular}{c|cccccc|c}
      $m$ & $q$ & $i$ & $\Omega$ & $\omega$ & $L$
      & $\sin B$
      & Case\\
      \hline
      \hline
      0 & $q_i$ & $i_i$ &$\Omega_i$ & $\omega_i$
      & $L_i$
      & $\sin B_i$ & circulation/libration\\
      \hline
      odd & $q_i$ & $i_i$ & $\Omega_i-\pi$ & $\omega_i+\pi$
      & $L_i$
      & $-\sin B_i$ 
      & circulation\\
      \cline{5-8}
      & & & & $\omega_i$&$L_i-\pi$
      & $\sin B_i$       
      & libration\\
      \hline
      even & $q_i$ & $i_i$ & $\Omega_i$ & $\omega_i$& $L_i$
      & $\sin B_i$ & circulation/libration\\
      \hline
    \end{tabular}
    \caption{Orbital Elements for $q\simeq q_i$, i.e., at $t=mP_\chi$
      under the Quasi-rectilinear Approximation}
    \label{tb:1}
  \end{center}
\end{table}

\begin{figure}[hbtp]
  \begin{center}
   \resizebox{15cm}{!}{\includegraphics{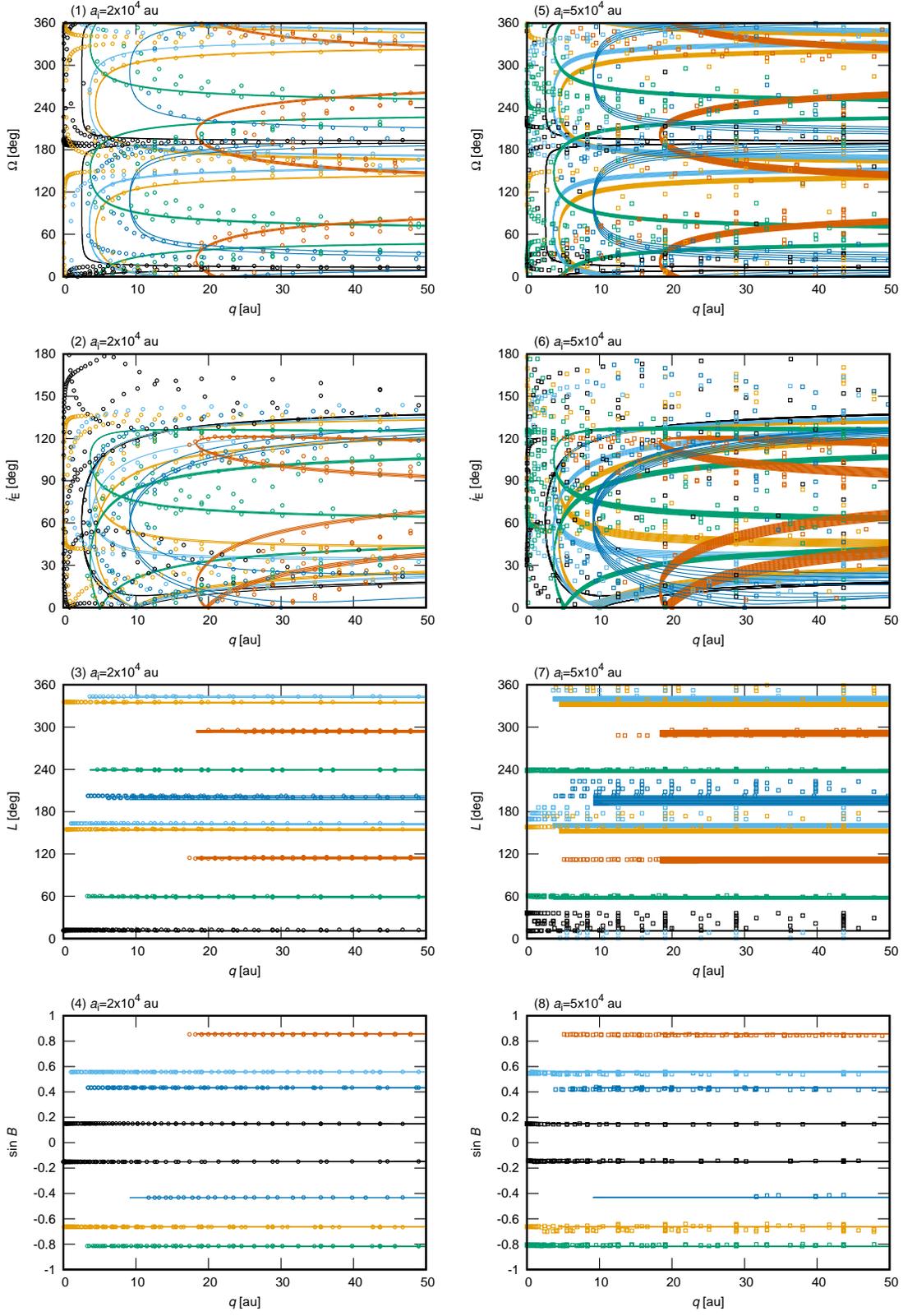}}
   \caption{
     Evolution of $\Omega$, $i_{\rm E}$, $L$, and $\sin B$
     of bodies orbiting around the Sun with
     the tidal forces from the Galactic disk
     plotted against $q<50$ au.
     Circles/squares are obtained by numerical integration of
     Equation (\ref{eq:em}) and the solid curves are analytical solutions.
     Left and right panels are for bodies with $a_i=2\times10^4$ au
     (circles) and $a_i=5\times10^4$ au (squares), respectively.
     Other initial conditions and the meaning of colors are the
     same as in Figure \ref{fig:t_oe1}.
     The output intervals of the results of the numerical calculations
     are not even in time in this figure.
    }
    \label{fig:q_oe}
  \end{center}
\end{figure}

Figure \ref{fig:q_oe} shows the analytical solutions to $\Omega$, $i_E$, $L$,
and $\sin B$ against $q<50$ au
with the results of numerical calculations for 4.5 Gyr
presented in Section \ref{ss:num}.
Panel (1) in Figure \ref{fig:q_oe} shows the behavior of $\Omega$
for $a_i=2\times10^4$ au.
As seen from Equation (\ref{eq:dlodt}), the analytical solution
to $\Omega$ (solid curves) drastically changes when $q$ is
near its minimum.
For example, for a body with $\omega_i=10^\circ$ (black),
$\Omega$ can have almost any value between 0 and 360$^\circ$ when $q$ is small.
The curves overlap every two oscillations of $q$
since $P_{\Omega^*}/P_\chi\simeq 2$.
The disagreements between the analytical solution and
the numerical calculations (circles) are not negligible
in the observable region, i.e., $q\lesssim 10$ au.
Panel (2) in Figure \ref{fig:q_oe} shows the behavior of $i_{\rm E}$
for $a_i=2\times10^4$ au.
Since $i_{\rm E}$ is a function of $i$ and $\Omega$ as
\begin{equation}
  \cos i_{\rm E} = \cos i\cos i_\odot+\sin i\sin i_\odot\cos\Omega,
\end{equation}
the prediction of $i_{\rm E}$ in the observable region
using the analytical solution
is as difficult as much as that for $i$ (see Figure \ref{fig:iq})
and as sensitive to $q$ as much as that for $\Omega$.
Panels (5) and (6) in Figure \ref{fig:q_oe} are the same as
panels (1) and (2) but for $a_i=5\times10^4$ au.
The drift of the curves is seen more obviously than in panel (1)
since bodies with $a_i=5\times10^4$ au make more
oscillations in 4.5 Gyr.

Panels (3) and (4) in Figure \ref{fig:q_oe} show the behavior of
$L$ and $\sin B$ for $a_i=2\times10^4$ au, respectively, 
and panels (7) and (8) in Figure \ref{fig:q_oe} are the same as panels (3) and (4),
respectively, but for $a_i=5\times10^4$ au.
Both $L$ and $\sin B$ are almost independent of $q$ for $q<50$ au.
The agreement of the results of numerical calculations and
the analytical solutions is good for $a_i=2\times10^4$ au.
For $a_i=5\times10^4$ au, the disagreement is larger than $\sim 30^\circ$
for some bodies beyond 4.5 Gyr but still much better than that
in $\Omega$ or $i_{\rm E}$.
Therefore, we conclude that the relation among $q$, $L$, and $\sin B$
in Table \ref{tb:1}, for any $m$,
is safely satisfied for $q$ in the observable region.

\subsection{The Empty Ecliptic}
Based on the standard scenario of the formation of the Oort cloud,
the initial orbital elements
of the Oort cloud comets are restricted as follows;
$q_i\lesssim 30$ au to be near a giant planet,
$i_i\simeq i_\odot=60^\circ$ and $\Omega_i\simeq\Omega_\odot=186^\circ$
to be on the ecliptic plane.
Also $\omega_i$ is uniformly distributed for $0\le\omega_i< 360^\circ$,
$L_i$ is given by Equation (\ref{eq:elu}),
and $\sin B_i$, in order to be on the ecliptic plane, is given by
\begin{equation}
  \tan B_i=\sin (L_i-\Omega_\odot)\tan i_\odot.
  \label{eq:be}
\end{equation}
The relation between $L$ and $\sin B$ in Table \ref{tb:1}
defines two planes in the Galactic coordinates.
As $L_i$ and $B_i$ are assumed to be on the ecliptic plane
the points that satisfy $0\le L<360^\circ$ and $\sin B$
given by Equation (\ref{eq:be}) for $L$ draw a curve on
the $L-\sin B$ plane, by definition.
Another set of points for an odd $m$ that satisfy
$0\le L<360^\circ$ and $-\sin B$
draw a curve that defines a second plane,
which is formed by a rotation 
of the ecliptic around the Galactic pole by 180$^\circ$:
\begin{equation}
  \tan B=-\sin (L-\Omega_\odot)\tan i_\odot.
  \label{eq:bee}
\end{equation}
We call this plane as ``the empty ecliptic,''
since it is not initially populated
and this plane and the ecliptic are symmetrical
about the plane perpendicular to the Galactic plane
through the intersection of the ecliptic plane
and the Galactic plane
(just like the focus and the empty focus of an ellipse).
If $L$ and $B$ of long-period comets in the observable region
are concentrated on these two planes,
it would constitute observational evidence
that the comets were on the ecliptic plane at $t=0$.
Comets with relatively small values such as $a\sim 10^4$ au, which
satisfy $P_\chi=$4.5 Gyr (i.e., $m=1$ for present), are predicted
to be on the empty ecliptic for their
first return to the planetary region.

%\newpage
\section{Observational data}\label{ss:obs}
\begin{figure}[hbtp]
  \begin{center}
    \rotatebox{-90}{
   \resizebox{13cm}{!}{\includegraphics{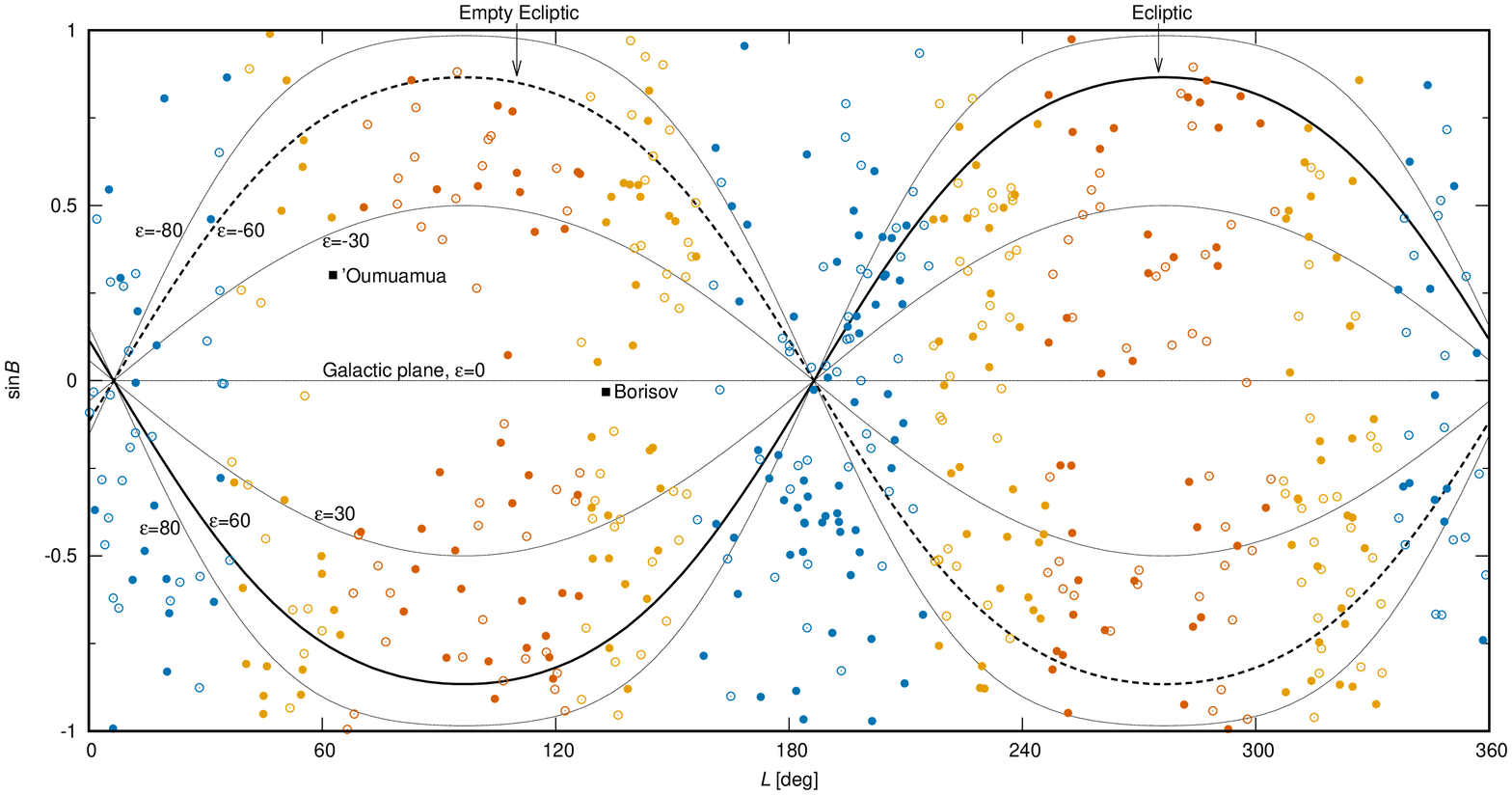}}}
    \caption{
      Solar system bodies with $q>1$ au and 
      $a>10^3$ au (open circles) or $e>1$ (filled circles)
      taken from
      the JPL Small Body Database Search Engine plotted
      on the $L-\sin B$ plane.
      Colors indicate the regions defined by $L'$ as
      $-30^\circ<L'<30^\circ$, $150^\circ<L'<210^\circ$ (blue),
      $30^\circ<L'<60^\circ$, $120^\circ<L'<150^\circ$,
      $210^\circ<L'<240^\circ$, $300^\circ<L'<330^\circ$,(orange),
      and
      $60^\circ<L'<120^\circ$, $240^\circ<L'<300^\circ$ (dark orange).
      Two interstellar objects, 1I/2017 U1 ('Oumuamua) and 2I/2019 Q4 (Borisov),
      are shown with squares.}
    \label{fig:lb}
  \end{center}
\end{figure}

Figure \ref{fig:lb} shows solar system bodies with
$q>1$ au and $a>10^3$ au or $e>1$
taken from the JPL Small Body Database Search Engine
on 2020 June 5
\footnote{https://ssd.jpl.nasa.gov/\_query.cgi}
on the $L-\sin B$ plane.
277 bodies with $e\le 1$ and 296 bodies with $e>1$ are
indicated with open and filled circles, respectively.
The bodies are divided into three groups with $L$
as indicated by colors.
Two interstellar objects, 1I/2017 U1 ('Oumuamua) and 2I/2019 Q4 (Borisov),
are additionally shown with squares for reference.
We used the osculating orbital elements to calculate $L$ and $B$
using Equations (\ref{eq:elu}) and (\ref{eq:sinb}).
To calculate $L$ and $B$ for bodies with $e>1$,
we replace $\omega$ in Equations (\ref{eq:elu}) and (\ref{eq:sinb}) as 
\begin{equation}
  \omega\rightarrow\omega +(\pi-f_\infty),
\end{equation}
where
\begin{equation}
  \cos f_\infty=-\frac{1}{e}.
\end{equation}
To evaluate the concentrations on the two planes,
we define the new angle $\varepsilon$ as
\begin{equation}
  \tan \varepsilon = \frac{\tan B}{\sin (L-\Omega_\odot)}.
  \label{eq:epsilon}
\end{equation}
The angle $\varepsilon$ is interpreted as a longitude around
the intersection of the ecliptic and the Galactic plane.
For the ecliptic and empty ecliptic planes,
$\varepsilon=i_\odot=60^\circ$ and $\varepsilon=-i_\odot=-60^\circ$,
respectively.
The solid and dashed curves in Figure \ref{fig:lb} show
the ecliptic and empty ecliptic planes, respectively.
Curves for $\varepsilon=0$, $\pm 30^\circ$, and $\pm 80^\circ$ are
also shown as thin dashed curves in Figure \ref{fig:lb}.

\begin{figure}[hbtp]
  \begin{center}
   \resizebox{7cm}{!}{\includegraphics{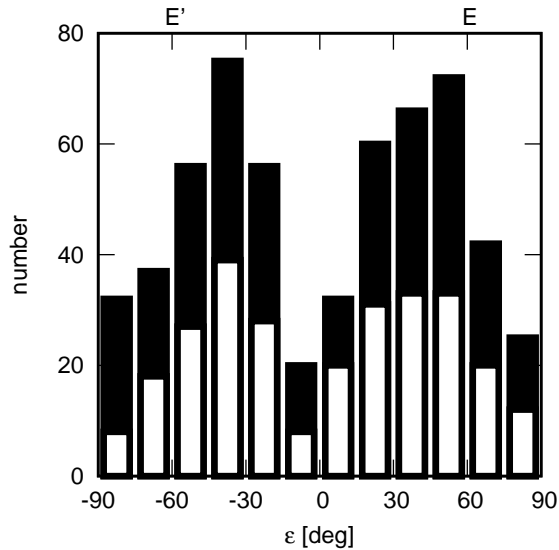}}
   \caption{
     Distribution of $\varepsilon$ defined by Equation (\ref{eq:epsilon})
     for all bodies in Figure \ref{fig:lb} except 'Oumuamua and Borisov.
     Filled bars show bodies with $e>1$.
     $E$ and $E'$ denote the ecliptic ($\varepsilon=60^\circ$) and
     the empty ecliptic ($\varepsilon=-60^\circ$), respectively.
   }
    \label{fig:bu_epsilon}
  \end{center}
\end{figure}

Figure \ref{fig:bu_epsilon} shows the distribution of $\varepsilon$.
There are two sharp peaks not exactly at the
ecliptic or empty ecliptic plane but near them.
An isotropic distribution would be flat in $\varepsilon$.

\begin{figure}[hbtp]
  \begin{center}
   \resizebox{15cm}{!}{\includegraphics{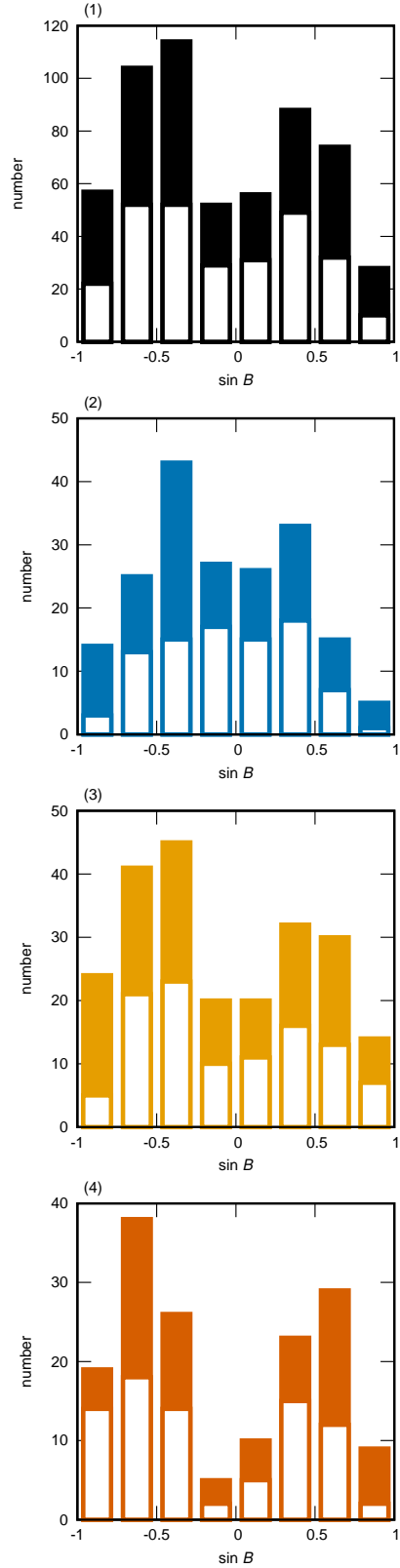}}
   \caption{
     Distribution of $\sin B$ for all bodies in Figure \ref{fig:lb}
     except 'Oumuamua and Borisov (panel (1)),
     for regions of $L$:
     blue (panel (2), $-30^\circ<L'<30^\circ$, $150^\circ<L'<210^\circ$),
     orange (panel (3), $30^\circ<L'<60^\circ$, $120^\circ<L'<150^\circ$,
     $210^\circ<L'<240^\circ$, $300^\circ<L'<330^\circ$),
     and dark orange (panel (4), $60^\circ<L'<120^\circ$, $240^\circ<L'<300^\circ$).
     Filled bars show bodies with $e>1$.
    }
    \label{fig:bu_sb}
  \end{center}
\end{figure}

Panel (1) in Figure \ref{fig:bu_sb} shows the distribution of $\sin B$.
The depletions around $b=0$ and $b=\pm 90^\circ$ found by
\citet{1984A&A...141...94L} and \citet{1987A&A...187..913D} 
are seen (note that $b=-B$); however, the shape of the distribution
depends on the regions of $L'=L-\Omega_\odot$.
Panels (2)-(4) in Figure \ref{fig:bu_sb} show
the distribution of $\sin B$ for regions of $L$:
blue ($-30^\circ<L'<30^\circ$, $150^\circ<L'<210^\circ$),
orange ($30^\circ<L'<60^\circ$, $120^\circ<L'<150^\circ$,
$210^\circ<L'<240^\circ$, $300^\circ<L'<330^\circ$),
and dark orange ($60^\circ<L'<120^\circ$, $240^\circ<L'<300^\circ$).
The less-sharp, rather a broad peak at $|\sin B|\le 0.5$ in panel (2) 
and
the sharpest double peaks at $|\sin B|>0.5$ in panel (4)
are explained as a consequence of the double peaks in
the distribution of $\varepsilon$.
If the depletions are
the result of the strength of the Galactic tide
as \citet{1987A&A...187..913D} explained,
the distributions are expected to be independent of $L'$
since the strength of the Galactic tide is independent of $L$. 
Therefore, we conclude that the concentration of comets
on the ecliptic and empty ecliptic planes
is a better explanation than that by \citet{1987A&A...187..913D}.

%\newpage
\section{Summary and Discussion}
We derived analytical solutions for $L$ and $B$,
the Galactic longitude and latitude of the aphelion direction
of bodies orbiting around the Sun, with the perturbation
from the Galactic disk in an axisymmetric approximation.
We used the solutions to predict the distribution of
observed long-period comets in the Galactic coordinates.
To evaluate the analytical solutions, we performed numerical
calculations of the orbital evolution including the effect
of the radial component of the Galactic tide,
which is neglected in the derivation of the analytical solutions.
Our findings are summarized as follows.
\begin{enumerate}
\item
  For bodies initially having eccentricities $e\simeq 1$,
  the analytical solutions and the results of numerical calculations
  show good agreement
  in the time evolutions normalized by the periods of the evolution.
  However, the Galactic inclination $i$ and
  the longitude of the ascending node in the Galactic coordinates $\Omega$
  show non-negligible disagreement,
  especially when their perihelion distances $q$
  are small enough to be in the observable region
  since the vertical angular momenta of the bodies 
  are not completely conserved.
  
\item In the orbital evolution, three periods are defined:
  (1) $P_\chi$, the period of oscillation of $e$
  (i.e., $q$), $i$, and $B$,
  (2) $P_{\Omega^*}$, the mean period of circulation of $\Omega$,
  and (3) $P_{L^*}$, the mean period of circulation of $L$.
  The period of the argument of perihelion, $\omega$,
  is $P_\chi$ for the case of libration (i.e., von Zeipel-Lidov-Kozai mechanism)
  and 2$P_\chi$ for the case of circulation.
  Under the quasi-rectilinear approximation (i.e., $e_i\sim 1$),
  the following relations are established among the analytical solutions;
  $P_{\Omega^*}/P_\chi\simeq 2$ for all cases
  and $P_{L^*}/P_\chi\simeq 2$ and $P_{L^*}\sim \infty$ 
  for the cases of libration 
  and circulation of $\omega$, respectively.
  Consequently, the evolutions of $q$ and $\Omega$ are coupled in any case,
  the evolutions of $q$ and $L$ are coupled in the case of libration of $\omega$,
  and $L$ evolves very little in the case of circulation of $\omega$.
  
\item
  Under the quasi-rectilinear approximation,
  the coupled evolutions of $L$ and $B$ of bodies initially on the
  ecliptic plane with $q$ in the planetary region
  draw two curves on the $L-B$ plane when their $q$ are small.
  One corresponds to the ecliptic plane and the other to
  the empty ecliptic defined by 
  the longitude around the intersection of the ecliptic and
  the Galactic plane $\varepsilon$,
  (see Equation (\ref{eq:epsilon})),
  $\varepsilon=60^\circ$ and $-60^\circ$, respectively.
  The numerical calculations showed that the coupling of $L$ and $B$
  is quite stable at any $q$ in the observable region
  and confirmed that $\varepsilon$ would be a reliable indicator of
  the dynamical character of observed long-period comets.
  The evolution of $\Omega$ is also coupled with that of $q$, $i$,
  $\omega$, and $B$ under the rectilinear approximation;
  however, $\varepsilon$ is a better indicator than $\Omega$ and others
  since the time variation of $\Omega$ is quite large at small $q$ and
  the value of $i$ is not nicely reproduced by the analytical solution
  at small $q$.
  
\item The distribution of $\varepsilon$ of observed solar system bodies
  with $q>1$ au and the semimajor axis $a>10^3$ au or $e>1$
  shows the double peaks that might correspond to the ecliptic 
  and empty ecliptic planes although their locations are not exactly at
  $\varepsilon=\pm 60^\circ$.
  The concentration of the bodies on the ecliptic and empty ecliptic planes
  explains the depletions around $B=0$ and $B=\pm 90^\circ$ 
  \citep{1984A&A...141...94L,1987A&A...187..913D}
  better than the explanation by \citet{1987A&A...187..913D}.
\end{enumerate}

The concentration of long-period comets from the Oort cloud
on the ecliptic and empty ecliptic planes is an observational
evidence that the Oort cloud comets were planetesimals initially
on the ecliptic plane.
We expect the concentrations even when we consider the effect
of passing stars.
Perturbations from passing stars change the conserved quantities
and may break the relation between $q$, $B$, and $L$ more or less;
however, it takes a much longer time to change the eccentricity
vector (i.e., $L$ and $B$) than to change $i$ \citep{2015AJ....150...26H}.
Therefore, we suggest that observers, including the space mission
Comet Interceptor, focus on the ecliptic plane and/or the
empty ecliptic plane to find dynamically new comets.

What we showed in Section \ref{ss:obs} is a brief examination.
An investigation of the distribution of observed small bodies 
has to include many factors.
The bodies should be carefully chosen from the database
and examined by classes defined by their
original semimajor axes calculated with non-gravitational
forces for active comets \citep[e.g.,][]{2014A&A...571A..63K}.
The orbital elements during the last perihelion passage would be
a key to the dynamical evolution if they encountered any of the planets
\citep{2009Sci...325.1234K, 2018A&A...620A..45F}.
Comparison with numerical calculations with all perturbations 
from the Galactic disk, stars, and planets is also important.
The long-term behavior found in numerical calculations
of comets in \citet{2020CeMDA.F} is the one that describes the
empty ecliptic plane.
Observational bias should also be taken into account.
Detailed examination of the distribution of long-period comets will
be our future work.
The all-sky survey by the Large Synoptic Survey Telescope
will provide valuable information for this study.

%% Putting eqnarrays or equations inside the mathletters environment groups
%% the enclosed equations by letter. For instance, the eqnarray below, instead
%% of being numbered, say, (4) and (5), would be numbered (4a) and (4b).
%% LaTeX the paper and look at the output to see the results.

%\newpage

\acknowledgments

I am grateful to Melaine Saillenfest and Takashi Ito
for their comments that greatly improved the quality of this paper
and to Marc Fouchard and Eiichiro Kokubo for their comments
and discussion that led to this work.
I also thank David Jewitt for carefully reading the manuscript.
Finally, I thank Giovanni B. Valsecchi for reviewing this paper. 
The numerical computations were in part carried out
on PC cluster at Center for Computational Astrophysics,
National Astronomical Observatory of Japan.
This work was partially supported by
the Programme National de Planetologie (PNP) of CNRS/INSU,
co-funded by CNES.





\appendix
\section{The initial Inclination given by planetary scattering}

Assume a planet in a circular orbit with the semimajor axis $a_{\rm planet}$
and a body with the inclination
with respect to the orbital plane of the planet $i$.
The relative velocity between the unperturbed orbits of
the planet and the comet $\Delta{\bf v}$ is
is written as the two components
\begin{equation}
  \Delta v_z=v_{\rm planet}\sqrt{3-T}\sin i, \quad
  \Delta v_r=v_{\rm planet}\sqrt{3-T}\cos i, 
\end{equation}
where $\Delta v_z$ is the component perpendicular to the orbital plane of the body,
$\Delta v_r$ is defined as $\Delta v_r=\sqrt{(\Delta v)^2-(\Delta v_z)^2}$,
$v_{\rm planet}$ is the velocity of the planet, and
$T$ is the Tisserand parameter with respect to the planet defined by
\begin{equation}
  T=\frac{a_{\rm planet}}{a}+2\sqrt{\frac{a}{a_{\rm planet}}(1-e^2)}\cos i,
\end{equation}
where $a$ and $e$ are the semimajor axis and eccentricity of the body.
Under the two-body approximation, 
the velocity of the body after planetary scattering ${\bf v}$ is
$v=v_{\rm planet}+\Delta v$
at maximum and
$v=v_{\rm planet}-\Delta v$
at minimum for each component.
Therefore, the maximum change in inclination $\Delta i$
given by planetary scattering
is approximated as a function of $T$ and $i$,
\begin{equation}
  \tan \Delta i \simeq
  \frac{\Delta v_z}{v_{\rm planet}-\Delta v_r}
  \simeq\frac{\sqrt{3-T}\sin i}{1-\sqrt{3-T}\cos i}.
  \label{eq:tandi}
\end{equation}
Bodies that we are interested in are those that
have $a'\gg a_{\rm planet}$,
where $a'$ is the semimajor axis after planetary scattering.
The value of $T$ for these bodies is estimated as follows.
To gain a change in velocity large enough to become nearly parabolic,
a large $\Delta v$ (i.e., a small $T$) is required.
For example, for a parabolic or hyperbolic orbit,
$\Delta v$ needs to satisfy
$|{\bf v}_{\rm planet}+\Delta{\bf v}|\ge v_{\rm esc}$,
where $v_{\rm esc}=\sqrt{2}v_{\rm planet}$ is the escape velocity at
the heliocentric distance $r=a_{\rm planet}$.
This condition leads to $T\le 2\sqrt{2}$.
On the other hand, the chance of having an effective encounter with
a planet becomes larger for smaller $\Delta v$ (i.e., for larger $T$)
since the gravitational radius of the planet is proportional to $\Delta v^{-2}$.
\citet{2006AJ....131.1119H} numerically showed that the efficiency of
having large $a'$ is higher for smaller $\Delta v$.
Therefore, we estimate that $T\simeq 2\sqrt{2}$ would be
the favored value for $a'\gg a_{\rm planet}$.
Substituting $T\simeq 2\sqrt{2}$ and $i\ll 1$ into Equation (\ref{eq:tandi}),
we have $\Delta i\ll 1$.  
More general and detailed discussion about post-encounter inclination
will be given by Valsecchi, G. B. et al. (2020, in preparation.)


\bibliography{empty_bib}{}
%\bibliography{sample63}{}
\bibliographystyle{aasjournal}



%%\listofchanges
\end{document}